\documentclass[12pt]{article}
\columnwidth\textwidth

\begin{document}

\newcommand{\be}{\begin{equation}}
\newcommand{\ee}{\end{equation}}
\newcommand{\beann}{\begin{eqnarray*}}
\newcommand{\eeann}{\end{eqnarray*}}
\newcommand{\bea}{\begin{eqnarray}}
\newcommand{\eea}{\end{eqnarray}}
\newcommand{\nn}{\nonumber}
\newcommand{\ben}{\begin{enumerate}}
\newcommand{\een}{\end{enumerate}}
\newtheorem{df}{Definition}
\newtheorem{thm}{Theorem}
\newtheorem{lem}{Lemma}
\newtheorem{prop}{Proposition}
\begin{titlepage}

\noindent
\begin{center}
{\LARGE Spherically symmetric gravitating shell as a reparametrization
  invariant system}

\vspace{2cm}

P.~H\'{a}j\'{\i}\v{c}ek \\
Institute for Theoretical Physics \\
University of Bern \\
Sidlerstrasse 5, CH-3012 Bern, Switzerland \\

\vspace*{2cm}

August 1997 \\ \vspace*{1cm}

\nopagebreak[4]

\begin{abstract}
  The subject of this paper are spherically symmetric thin shells made of
  ba\-ro\-trop\-ic ideal fluid and moving under the influence of their own
  gravitational field as well as that of a central black hole; the
  cosmological constant is assumed to be zero. The general super-Hamiltonian
  derived in a previous paper is rewritten for this spherically symmetric
  special case. The dependence of the resulting action on the gravitational
  variables is trivialized by a transformation due to Kucha\v{r}. The
  resulting variational principle depends only on shell variables, is
  reparametrization invariant, and includes both first- and second-class
  constraints. Several equivalent forms of the constrained system are written
  down. Exclusion of the second-class constraints leads to a super-Hamiltonian
  which appears to overlap with that by Ansoldi et al.\ in a quarter of the
  phase space. As Kucha\v{r}' variables are singular at the horizons of both
  Schwarzschild spacetimes inside and outside the shell, the dynamics is first
  well-defined only inside of 16 disjoint sectors. The 16 sectors are,
  however, shown to be contained in a single, connected symplectic manifold
  and the constraints are extended to this manifold by continuity. Poisson
  bracket between no two independent spacetime coordinates of the shell vanish
  at any intersection of two horizons.

\end{abstract}

\end{center}

\end{titlepage}

\section{Introduction}
\label{sec:intro}
Spherically symmetric thin shells are popular models used extensively in the
study of a number of phenomena: properties of classical gravitational collapse
\cite{israel-coll}, properties of classical black holes \cite{farrug}, quantum
gravitational collapse \cite{H-K-K}, the dynamics of domain walls in early
Universe \cite{guth} and \cite{A-A-B-S}, the back reaction in Hawking effect
\cite{K-W}, entropy on black holes \cite{york} or quantum theory of black
holes \cite{berez}, to mention just few examples.

Attempts to derive a Hamiltonian formalism for such shells are for example
Refs.\ \cite{B-K-T}, \cite{K-W}, \cite{L-F-W}, \cite{H-B}, \cite{A-A-B-S} and
\cite{H-K}. The Hamiltonian (or super-Hamiltonian) is either guessed directly
from equations of motion (Refs.\ \cite{B-K-T} and \cite{H-B}), or it is
derived from a variational principle guessed for the spherically symmetric
system consisting of dust shells and gravity (Refs.\ \cite{K-W} and
\cite{L-F-W}), or it is derived from the Lagrangian formalism based on the sum
of the Einstein-Hilbert action and an action for ideal fluid either after
reducing the action by spherical symmetry \cite{A-A-B-S} or without any
assumption about symmetry \cite{H-K}.

In Ref.\ \cite{H-K}, both super-Hamiltonian and the symplectic structure are
derived from the Einstein-Hilbert-ideal-fluid variational principle. In this
sense, the symplectic structure is unique; it contains a boundary term at
the hypersurface of the shell and it turns out that the momentum conjugate to
the surface area of the shell is the (hyperbolic) angle between the shell and
the foliation hypersurface (`Kijowski momentum', \cite{JK2}; cf.\ also
Ref.\ \cite{H-H}). This momentum will play an important role in our
calculations. 

In the present paper, we shall derive a super-Hamiltonian and a symplectic
form for the spherically symmetric ideal fluid shells, starting from the
general formula of Ref.\ \cite{H-K}. For the sake of simplicity, we shall also
assume that the cosmological constant and all fields different from gravity
are zero. Our leading principle is the reparametrization invariance.  Thus,
the result must be a super-Hamiltonian rather than a Hamiltonian. One problem
is then how the variables describing the gravitational field around the shell
can be made to disappear from the action so that the final formalism contains
the shell variables only. As most of these gravitational variables just
describe a gauge, one possible method is to choose a gauge and to reduce the
system, as for instance in Refs.\ \cite{L-F-W} and \cite{H-K2}; then, however,
the reparametrization invariance is lost. We find a suitable tool in a
transformation due to Kucha\v{r} \cite{kuch1}. This transformation trivializes
the gravitational part of the equations of motion to such an extent that they
do not contain any more information about the motion of the shell. The
boundary terms that result from Kucha\v{r}' transformation contribute to the
shell part of the symplectic form. They not only modify Kijowski's momentum
but provide additional terms so that this part itself becomes non-degenerate;
thus, the symplectic structure of the shell emerges. Several equivalent forms
of the variational principle can be written down.

For example, one of the resulting phase spaces is locally described by four
pairs of conjugate quantities, namely $(E_+,T_+)$, $(E_-,T_-)$, $(P_+,R_+)$
and $(P_-,R_-)$, where $T_\pm$ and $R_\pm$ are the Schwarzschild coordinates,
$E_\pm$ the Schwarzschild masses, and $P_\pm$ the modified Kijowski's momenta;
the sign `$+$' refers to the outside and `$-$' to the inside Schwarzchild
spacetimes. There are then three constraints: 1) the super-Hamiltonian
constraint $C_s = 0$ is (roughly) the time-time component of Israel's matching
condition at the shell and it is a primary constraint, 2) the continuity
condition $R_+ - R_- = 0$ is another primary and 3) the Poisson bracket $\chi
:= \{C_s,R_+ - R_-\}$ fails to vanish, so $\chi = 0$ is a secondary
constraint. The two constraint functions $\chi$ and $R_+ - R_-$ form a
second-class pair. The second class constraints can be solved for $[R]$ and
$\bar{P}$ \footnote{We adhere to the usual notation in the theory of thin
  shells: for any, possibly discontinuous, function $X$ at the shell, $\bar{X}
  := (X_+ + X_-)/2$ and $[X] := X_+ - X_-$, where $X_\pm$ are the limits of
  $X$ at the shell, $X_+$ from right and $X_-$ from left.}, and the solution
can be substituted back into the action; in this manner, a partially reduced
system with three pairs of conjugate variables $(E_+,T_+)$, $(E_-,T_-)$ and
$([P],\bar{R})$ and just one constraint $C_s^r = 0$ is obtained. In four
sectors of the phase space, $C_s^r$ has a similar form as the
super-Hamiltonian of Ref.\ \cite{A-A-B-S}, which has been derived in a
completely different way. The origin of the second-class constraints is in the
additional conditions by which the general Einstein-Hilbert-ideal-fluid action
must be supplemented in order that the system with a shell be well-defined:
the so-called continuity conditions (see Ref.\ \cite{H-K} and the next
section).

The Schwarzschild coordinates $(T_\pm,R_\pm)$ of the shell are singular at the
horizons of the spacetimes inside and outside the shell. Each of these two
spacetimes is separated by the horizons into four quadrants.  As a consequence
of this, the phase space of the system is split up into 16 disjoint sectors.
The dynamical trajectories that result from the action can be smoothly matched
through the horizons, because the shells are regular there. This suggests that
there are dynamical variables which are regular at the horizons; we try
Eddington-Finkelstein's and Kruskal's transformations. The first one leads to
an atlas of 16 overlapping Darboux charts covering a single, connected
extension of the old phase space; this extension is not maximal, however,
because the points at the intersections of the horizons are not covered. The
second transformation leads to one single chart covering the maximal extension
of the old phase space. The constraints have smooth extensions to new
phase space in both cases. The Poisson bracket between the Kruskal coordinates
$u$ and $v$ of the shell does not vanish and we show that this must be true for
{\em any} spacetime coordinates that are regular at the intersection of two
horizons. 

Our super-Hamiltonians are reparametrization invariant, but rather
complicated: they depend on momenta through exponentials and square roots.
Some problems arise immediately. For example, the problem of quantizing such
complicated super-Hamiltonians or the problem of relation between the
super-Hamiltonians of the present paper and that of Ref.\ \cite{H-B}, which is
not only reparametrization invariant but also quadratic in momenta. These
problems will not be addressed here.

The plan of the paper is as follows. In Sec.\ 2, we introduce the general
formula for super-Hamiltonian from Ref.\ \cite{H-K}. In this way, the paper
becomes self-contained. In Sec.\ 3, the assumption of spherical symmetry is
formulated, the dynamical variables are adapted to the symmetry, and the
action is expressed as a functional of these variables. In Sec.\ 4,
Kucha\v{r}' transformation is performed and an effective shell
super-Hamiltonian is derived. Sec.\ 5 is devoted to the study of the shell
action obtained in Sec.\ 4. We check that correct equations of motion result
from it, investigate the structure of the constrained system defined by the
action, and remove the second-class constraint by a partial reduction.
Finally, Sec.\ 6 addresses the problem of the singularity at the horizons.

We use the units such that c = G = 1 (c is the velocity of light in vacuum and
G is Newton's constant).

\section{The spacetime and the shell}
In this section, we describe the spacetime with the shell, introduce the basic
ideas and quantities and collect the equations from Ref.\ \cite{H-K} that will
be needed as a starting point of our investigation.

Let $({\mathcal M},g)$ be an assymptotically flat globally hyperbolic
spacetime and let a thin shell of ideal fluid move along a timelike
hypersurface $\Sigma$ in ${\mathcal M}$; $\Sigma$ divides the spacetime into
two parts, ${\mathcal M}_+$ and ${\mathcal M}_-$ so that ${\mathcal M}_+$ is
adjacent to the infinity where the observers are. Let $x_{(\pm)}^\mu$ be some
coordinates in ${\mathcal M}_\pm$ and $\xi^\alpha$ be some in $\Sigma$. No
relation between the coordinates $x_{(-)}^\mu$ and $x_{(+)}^\mu$ is assumed.
Let $x_{(\pm)}^\mu(\xi)$ be the embedding functions of $\Sigma$ in ${\mathcal
  M}_\pm$. We assume that (see Ref.\ \cite{H-K})
 \be 
   \gamma_{\alpha\beta}(\xi) =
   \left(g_{(-)\mu\nu}\frac{\partial x^\mu_{(-)}}{\partial
    \xi^\alpha}\frac{\partial x^\nu_{(-)}}{\partial
    \xi^\beta}(x_{(-)}(\xi))\right)_- = \left(g_{(+)\mu\nu}\frac{\partial
    x^\mu_{(+)}}{\partial \xi^\alpha}\frac{\partial x^\nu_{(+)}}{\partial
    \xi^\beta}(x_{(+)}(\xi))\right)_+,
\label{gamma-g}
\ee
where the symbols $()_\pm$ denote the limits from the four-volumes ${\mathcal
  M}_\pm$ towards $\Sigma$, $g_{(\pm)\mu\nu}$ is the metric in ${\mathcal
  M}_\pm$ with respect to the coordinates $x_{(\pm)}^\mu$ and
$\gamma_{\alpha\beta}(\xi)$ is the metric in $\Sigma$ with respect to
$\xi^\alpha$. Eqs.\ (\ref{gamma-g}) are called {\em continuity relations}.

Let $\{S_t\}$ be a foliation of ${\mathcal M}$ by Cauchy hupersurfaces $S_t$,
where $t$ runs through some real interval and let $S_{(\pm)t} := S_t \cap
{\mathcal M}_\pm$. We assume that $S_t$ are (continuous) hypersurfaces in $M$
and that $S_{(\pm)t}$ are smooth hypersurfaces in ${\mathcal M}_\pm$, for all
$t$. The ADM-like formalism described in Ref.\ \cite{H-K} is based on a choice
of coordinates $x_{(\pm)}^\mu$ that are adapted to the foliation $\{S_t\}$ on
one hand and to $\Sigma$ on the other. Such coordinates satisfy the following
requirements.  First,
\[
  x_{(\pm)}^0 = t,\quad \xi^0 = t;
\]
then, $x_{(\pm)}^k$, $k = 1,2,3$, can be considered as coordinates on $S_t$
and $\xi^K$, $K = 1,2$, as coordinates om $\Sigma \cap S_t$. Second, the
embedding functions $x_{(\pm)}^\mu(\xi)$ defining $\Sigma$ satisfy
\beann
  x_{(\pm)}^0(\xi^0,\xi^1,\xi^2) & = & \xi^0, \\
  \frac{d}{d\xi^0}x^k_{(\pm)}(\xi^0,\xi^1,\xi^2) & = & 0
\eeann
for all $(\xi^0,\xi^1,\xi^2) \in \Sigma$ and $k = 1,2,3$. Thus, the vector
$\partial/\partial t$ is tangential to $\Sigma$. The functions
\[
  y_{(\pm)}^k(\xi^1,\xi^2) := x^k_{(\pm)}(\xi^0,\xi^1,\xi^2)
\]
can be considered as embedding functions of the surface $\Sigma \cap S_t$ in
the hypersurfaces $S_{(\pm)t}$; they are independent of the time coordinate
$t$. Thus, the dynamics of the shell is completely determined by the time
dependence of the metric and of the matter fields along $\Sigma$. This leads
to a great simplification of the formalism and of the variational procedure,
but to no restriction of generality, see Ref.\ \cite{H-K} for a discussion of
this point. We shall often leave out the index $t$ in the sequel.

The 3+1 decomposition of the metric $g_{(\pm)\mu\nu}$ can be described as
follows (see Ref.\ \cite{M-T-W}):
\beann
  g_{(\pm)}^{00} = -N_{(\pm)}^{-2},&& g_{(\pm)0k} = N_{(\pm)k}, \\
  g_{(\pm)kl} = q_{(\pm)kl},&& g_\pm = -N_\pm^2 q_\pm,
\eeann
where $N_\pm$ is the lapse and $N_{(\pm)k}$ the shift in $S_\pm$,
$q_{(\pm)kl}$ is the metric induced in $S_\pm$ by $g_{(\pm)\mu\nu}$, $g_\pm$
is the determinant of $g_{(\pm)\mu\nu}$ and $q_\pm$ that of
$q_{(\pm)kl}$. (We work with adapted coordinates.) The 2+1 decomposition of
the metric $\gamma_{\alpha\beta}$ is analogous:
\beann
  \gamma^{00} = -\nu^{-2},&& \gamma_{0K} = \nu_K, \\
  \lambda_{KL} = \gamma_{KL},&& \gamma = -\lambda\nu^2,
\eeann
where $\lambda_{KL}$ is the metric of the surface $\Sigma \cap S$ with respect
to the coordinates $\xi^K$ and $\lambda$ its determinant. The 2+1
decomposition of the continuity relations (\ref{gamma-g}) is
\bea
  \nu & = & \sqrt{N^2_\pm - (N^\bot_{(\pm)})^2},
\label{2;1} \\
  \nu_K & = & N_{(\pm)k} \frac{\partial y_{(\pm)}^k}{\partial\xi^K},
\label{2;2} \\
  \lambda_{KL} & = & q_{(\pm)kl}\frac{\partial y_{(\pm)
  }^k}{\partial\xi^K}\frac{\partial y_{(\pm)}^l}{\partial\xi^L}, 
\label{2;3}
\eea
where
\[
  N^\bot_{(\pm)} = N_{(\pm)k} m^k_{(\pm)}, 
\]
and $m^k_{(\pm)}$ is the unit normal vector to $\Sigma \cap S$ tangent to
$S_{(\pm)}$ and oriented from $S_{(-)}$ to $S_{(+)}$ (towards the observers).
This orientation will be often used, so we call it {\em right} orientation.

An important role is played by the (hyperbolic) angle $\alpha_\pm$ of the two
hypersurfaces $\Sigma$ and $S_{(\pm)}$ which is defined by
\be
  \sinh\alpha_\pm :=
  -g_{(\pm)\mu\nu}n_{(\pm)}^\mu\tilde{m}_{(\pm)}^\nu,
\label{alpha}
\ee
where $n_{(\pm)}^\mu$ is the future-oriented unit normal to
$S_{(\pm)}$ and $\tilde{m}_{(\pm)}^\nu$ is the right-oriented unit normal to
$\Sigma$ in ${\mathcal M}_\pm$. One easily proves that
\[
  N_\pm = \nu\cosh\alpha_\pm,\quad N^\bot_{(\pm)} = \nu\sinh\alpha_\pm.
\]
 
Another important quantity is the second fundamental form $l_{KL}$ of the
surface $\Sigma \cap S$ in $S_{(\pm)}$, which is defined by
\[
  l_{(\pm)KL} := m_{(\pm)k|l}\frac{\partial y_{(\pm)
  }^k}{\partial\xi^K}\frac{\partial y_{(\pm)}^l}{\partial\xi^L};
\]
here the symbol `$|$' denotes the covariant derivative associated with the
metric $q_{(\pm)kl}$ in $S_{(\pm)}$. We reserve `$;$' for the covariant
derivative defined by $g_{\mu\nu}$ in ${\mathcal M}$ and `$:$' for that by
$\gamma_{\alpha\beta}$ in $\Sigma$. The trace $l^{KL}\gamma_{KL}$ of $l_{KL}$
will be denoted by $l$.

The matter of the shell is assumed to be relativistic barotropic perfect
fluid. Its description follows the pattern given in Refs.\ \cite{K-S-G} and
\cite{H-K}; let us collect the relevant formulas.

The mass points of the fluid fill the so-called matter space $Z$ which is a
two dimensional manifold for a shell. The coordinates $z^A$, $A = 1,2$, in $Z$
can be thought of as Lagrangian coordinates of the fluid. The state of the
fluid is described by the `fields' $z^A(\xi)$. The matter space carries a
scalar density $h(z)$, which determines the mole or particle density of the
fluid in the matter space. The mole (particle) current $j^\alpha$ in $\Sigma$
is given by
\[
  j^\alpha = h\epsilon^{\alpha\beta\gamma}z^1_\beta z^2_\gamma,
\]
where we use the abbreviation
\[
  z^A_\alpha := \frac{\partial z^A}{\partial \xi^\alpha}.
\]
The current $j^\alpha$ is identically conserved, $j^\alpha_{\ :\alpha} =
0$. $j^\alpha$ defines the three-velocity $u^\alpha(\xi)$ and the rest mole
(particle) density $n(\xi)$ in $\Sigma$ by 
\[
  j^\alpha = \sqrt{|\gamma|}nu^\alpha,
\]
where $\gamma_{\alpha\beta}u^\alpha u^\beta = -1$.

The information about the consecutive relations of the fluid is
encoded in the quantity $e(n)$ that gives the energy per mole in the rest
frame of the fluid as a function of the mole density $n$. Then, the surface
tension $-p$ of the fluid is determined by (see Ref.\ \cite{K-S-G})
\[
  p = n^2\frac{de}{dn}.
\]
The dynamics of the fluid in the fixed background three-spacetime $\Sigma$ can
be derived from the Lagrangian
\[
  L_m = -\sqrt{|\gamma|}\rho(n),
\]
where $\rho := ne(n)$ denotes the rest mass density of the fluid. The
stress-energy tensor density
\be
  T^{\alpha\beta} = \sqrt{|\gamma|}\left((\rho + p)u^\alpha u^\beta +
    p\gamma^{\alpha\beta}\right)
\label{1.6}
\ee
satisfies the relation
\be
  T^{\alpha\beta}(x) = 2\frac{\delta I_m}{\delta \gamma_{\alpha\beta}(x)},
\label{1.8}
\ee
where $I_m$ is the action of the fluid,
\[
  I_m = \int_\Sigma d^3\xi L_m.
\]
It also satisfies (the Noether identity)
\be
  T^\alpha_\beta = L_m\delta^\alpha_\beta - \frac{\partial L_m}{\partial
  z^A_\alpha}z^A_\beta.
\label{1.10}
\ee
The momenta $p_A$ of the fluid are defined by
\[
  p_A := \frac{\partial L_m}{\partial z^A_0}.
\]

The negative component $- T^0_0$ of the stress-energy tensor in the adapted
coordinates $\xi^\alpha$ is the Hamiltonian of the fluid \cite{H-K}. In
Ref.\ \cite{H-K}, the following important formulas have been derived: 
\be
  T^0_{0} = - \nu \sqrt{\lambda} \tilde{T}^{\bot\bot} - \nu^K \sqrt{\lambda}
  \tilde{T}^\bot_K 
\label{T00}
\ee
and
\be
  \frac{\partial T^0_{0}}{\partial\lambda_{KL}} = \frac{1}{2}T^{KL},
\label{Tlambda}
\ee
where
\bea
  \tilde{T}^{\bot\bot} & = & \frac{n}{\rho'(j^0)^2}
  \lambda^{KL}z^A_Kz^B_Lp_Ap_B + \rho, 
\label{Tbotbot} \\
  \tilde{T}^\bot_K & = & z^A_Kp_A.
\label{TbotK}
\eea
We have introduced the symbols
\[ 
  \tilde{T}^{\bot\bot} = 
  \frac{1}{\sqrt{|\gamma|}}T^{\alpha\beta}\tilde{n}_\alpha\tilde{n}_\beta,
  \quad \tilde{T}^\bot_{K} = 
  \frac{1}{\sqrt{|\gamma|}}T^\alpha_K\tilde{n}_\alpha,
\]
where $\tilde{n}^\alpha$ is the future-oriented unit normal to the surface
$\Sigma \cap S$ in $\Sigma$.

Finally, the master formula, the Hamiltonian of the whole system
consisting of the shell and gravity (the gravitational field being also
dynamical) reads (for a derivation, see Ref.\ \cite{H-K})
\bea
  \check{\mathcal H} & = & \int_{S_+} d^3x (N_+C + N_{(+)}^kC_k) + \int_{S_-}
  d^3x (N_-C + N_{(-)}^kC_k)\nn \\  
  & + & \int_{S\cap\Sigma}d^2\xi (\nu C_s + \nu^K C_{sK})
  + \frac{1}{8\pi}\int_{S\cap\Sigma^+}d^2\xi L^0_0,
\label{hamilt}
\eea
where $C$ is ADM super-Hamiltonian and $C_k$ is ADM supermomentum,
\[
  C = \frac{1}{16\pi}\left(
  \frac{2\pi^{kl}\pi_{kl} - \pi^2}{2\sqrt{q}} -
  \sqrt{q}R^{(3)}\right), 
\]
\[
  C_k = - \frac{1}{8\pi}\pi^l_{k|l},
\]
$\pi^{kl}$ the ADM-momentum for gravity and $R^{(3)}$ the curvature scalar of
the metric $q_{kl}$.  The surface super-Hamiltonian $C_s$ and the surface
supermomentum $C_{sK}$ at the shell are given by 
\beann 
  C_s & = &
  -\frac{1}{8\pi}[\tilde{\pi}^{\bot\bot}\sinh\alpha
  - l\cosh\alpha] + \tilde{T}^{\bot\bot}_s, \\
  C_{sK} & = &-\frac{1}{8\pi}[\tilde{\pi}^\bot_K + \alpha_{,K}] +
  \tilde{T}^\bot_{sK}, 
\eeann 
where
\[
  \tilde{\pi}^{\bot\bot} = \frac{\pi^{kl}}{\sqrt{q}}m_k m_l, \quad
  \tilde{\pi}^\bot_K = \frac{\pi^{kl}}{\sqrt{q}}q_{lr}m_k \frac{\partial
  y^r}{\partial\xi^K},\quad 
  \tilde{\pi}_{KL} = \frac{\pi^{kl}}{\sqrt{q}}\frac{\partial
  y^k}{\partial\xi^K}\frac{\partial y^l}{\partial\xi^L},
\]
The symplectic form is 
\bea
  \Omega(\delta\mathbf{X},\dot{\mathbf{X}}) 
  & = & \frac{1}{16\pi}\int_{S}d^3x\,
  (\delta\pi^{kl}\dot{q}_{kl} - \delta q_{kl}\dot{\pi}^{kl}) \nn \\
  & + & \frac{1}{16\pi}\int_{S\cap\Sigma}d^2\xi\
  (\delta[\alpha]\dot{\sqrt{\lambda}} - \delta\sqrt{\lambda}
    \dot{[\alpha]}) + \int_{S\cap\Sigma}d^2\xi\,
  (\delta p_A\dot{z}^A - \delta z^A\dot{p}_A) \nn \\
  & - & \frac{1}{16\pi}\int_{S\cap\Sigma^+}d^2\xi\,
  (\delta\alpha^+\dot{\sqrt{\lambda^+}} - \delta\sqrt{\lambda^+}
  \dot{\alpha}^+),
\label{varHcheck}
\eea
where the quantities with the upper index `$+$' concern the hypersurface
$\Sigma^+$ and
\beann
  \delta\mathbf{X} & = & \left(\delta\pi^{kl}(x),\delta q_{kl}(x),
  \delta[\alpha(\xi)], \delta\sqrt{\lambda}(\xi), \delta p_a(\xi), \delta
  z^a(\xi), \delta\sqrt{\lambda^+(\xi)}, \delta\alpha^+(\xi) \right), \\
  \dot{\mathbf{X}} & = & \left(\dot{\pi}^{kl}(x),\dot{q}_{kl}(x),
  [\dot{\alpha(\xi)}],\dot{\sqrt{\lambda(\xi)}}, \dot{p}_a(\xi),
  \dot{z}^a(\xi), \dot{\sqrt{\lambda^+(\xi)}}, \dot{\alpha}^+(\xi) \right) 
\eeann
are two vectors tangential to the symplectic manifold of the system.

The equations of motion follow from the variation formula (cf.\
Ref.\ \cite{H-K}) 
\be
  \delta\check{\mathcal H} = \Omega(\delta\mathbf{X},\dot{\mathbf{X}})
  + \frac{1}{16\pi}\int_{S\cap\Sigma^+}d^2\eta\,
  \gamma_{\alpha\beta} \delta Q^{\alpha\beta},
\label{varH}
\ee 
where $Q^{\alpha\beta} := L\gamma^{\alpha\beta} - L^{\alpha\beta}$ and $L
:= L^{\alpha\beta}\gamma_{\alpha\beta}$. This formula plays a double role. By
deriving it from the Lagrange formalism carefully considering all boundary
terms, we find what is the symplectic form of the system. By comparing the R.
H. S. and the L. H. S. coefficients at the variations of the same variable, we
obtain the equations of motion.

The last terms in Eqs.\ (\ref{hamilt}) and (\ref{varH}) determine the
so-called {\em control mode} (see Ref.\ \cite{K-T}). In fact, there must be
one such term for each infinity, see the next section. $\Sigma^+$
is a timelike surface that forms a boundary of $S$ and it will be pushed to
infinity eventually, $L_{\alpha\beta}$ is the second fundamental form of
$\Sigma^+$ defined by 
\[
  L_{\alpha\beta} := \tilde{m}_{\mu;\nu}\frac{\partial
  x^\mu}{\partial\xi^\alpha} \frac{\partial x^\nu}{\partial\xi^\beta},
\]
$\tilde{m}^\mu$ being the external (with respect to the volume closed by
$\Sigma^+$) unit normal to $\Sigma^+$ and $x^\mu(\xi)$ are the embedding
functions defining $\Sigma^+$. The usual canonical equations hold only if the
last term in Eq.\ (\ref{varH}) vanishes. This means that the field
$Q^{\alpha\beta}$ must be kept fixed at $\Sigma^+$. In Ref.\ \cite{JK2}, a
more natural control mode is described; it results, if we perform a Legendre
transformation from $\check{\mathcal H}$ to ${\mathcal H}$ by
\be
  {\mathcal H} = \check{\mathcal H} - \frac{1}{16\pi}
  \int_{\Sigma^+\cap S}d^2\xi\,\gamma_{KL}Q^{KL}
\label{mix-mode}
\ee
so that the boundary term in Eq.\ (\ref{varH}) becomes
\be
  \frac{1}{16\pi} \int_{\Sigma^+\cap S}d^2\xi\,(\gamma_{00}\delta
  Q^{00} + 2\gamma_{0K}\delta Q^{0K} - Q^{KL}\delta\gamma_{KL}).
\label{mix-bound}
\ee
If the surface $\Sigma^+$ is shifted to infinity and if the usual fall-off
conditions on $q_{kl}$, $\pi^{kl}$, $N$ and $N_k$ are met, then the on-shell
value of ${\mathcal H}$ is the ADM mass and the expression
(\ref{mix-bound}) vanishes (see Ref.\ \cite{JK2}). We will pass to this
description directly in the spherically symmetric case.

\section{Spherical symmetry}
In this section, we substitute the spherically symetric values of the physical
fields and foliation into the Hamiltonian (\ref{hamilt})
and the symplectic form (\ref{varHcheck}). We start with the transformation of
the volume terms following closely the notation by Kucha\v{r} \cite{kuch1}.

There are coordinates $t$, $r$, $\vartheta$ and $\varphi$ such that the
spacetime metric has the form
\[
  ds^2 = -(N^2 - N_r^2\Lambda^{-2})dt^2 + 2N_rdtdr + \Lambda^2dr^2 +
  R^2d\vartheta^2 +R^2\sin^2\vartheta d\varphi^2
\]
with the square root of the determinant
\[
  \sqrt{-g} = N\Lambda R^2\sin\vartheta,
\]
where $N(t,r)$, $N^r(t,r)$, $\Lambda(t,r)$ and $R(t,r)$ are some functions of
$t$ and $r$.  We assume that $r \in (-\infty,\infty)$, that $r = \pm\infty$
are spacelike infinities and that the equation $r = 0$ defines the shell. We
further assume that the coordinates are continuous across the shell. We shall
leave out the indices $\pm$, but we will keep in mind that some components of
the metric ($N$, $N_r$, $\Lambda$ etc.) are discontinuous across the shell.

The folii $t$ = const carry the metric $q_{kl}$:
\[
  ds^2 = \Lambda^2 dr^2 + R^2d\vartheta^2 +R^2\sin^2\vartheta d\varphi^2
\]
with the square root of the determinant
\[
  \sqrt{q} = \Lambda R^2\sin\vartheta.
\]
The shell hypersurface $\Sigma$ can be described by the coordinates $t$,
$\vartheta$ and $\varphi$ and the metric $\gamma_{\alpha\beta}$ satisfying the
continuity relations (\ref{gamma-g}) is
\[
  ds^2 = - (N^2 - N_r^2\Lambda^{-2})dt^2 + R^2d\vartheta^2 +R^2\sin^2\vartheta
  d\varphi^2.
\]

The components of the unit future-oriented vector $n$ normal to $S$ are $n_\mu
= - N\delta^0_\mu$. The corresponding second fundamental form $K_{kl}$ can
easily be calculated; its two independent components are
\beann
  K_{rr} & = & - \frac{\Lambda}{N}\left(\dot{\Lambda} - (\Lambda N^r)'\right),
\\
  K_{\vartheta\vartheta} & = & - \frac{R}{N}(\dot{R} - N^rR'),
\eeann
where the prime denotes the derivative with respect to $r$ and the dot that
with respect to $t$. Then, the ADM momentum $\pi^{kl}$ is determined by
\beann
  \pi^{rr} & = & - \frac{2R\sin\vartheta}{\Lambda N}(\dot{R} - N^rR'), \\
  \pi^{\vartheta\vartheta} & = & - \frac{\Lambda R^2\sin\vartheta}{N}
  \left(\frac{1}{\Lambda R^2}(\dot{\Lambda} - (N^r\Lambda)') +
  \frac{1}{R^3}(\dot{R}- N^rR')\right);
\eeann
We obtain for the Liouville form
\beann
  \lefteqn{\theta = \int drd\vartheta d\varphi\, \pi^{kl}dq_{kl} = } \\ 
  && -16\pi\int dr
  \left\{ \frac{R}{N} (\dot{R} - N^rR')d\Lambda +
  \left(\frac{R}{N}\left(\dot{\Lambda} - (N^r\Lambda)'\right) +
  \frac{\Lambda}{N}(\dot{R}-N^rR')\right)dR\right\}.
\eeann
Let us set, with Kucha\v{r}:
\beann
  P_\Lambda & = & - \frac{R}{N}(\dot{R} - N^rR'), \\
  P_R & = & -\frac{R}{N}(\dot{\Lambda} - (N^r\Lambda)') -
  \frac{\Lambda}{N}(\dot{R} - N^rR').
\eeann
Hence,
\[
  \pi^{rr} = \frac{2P_\Lambda}{\Lambda}\sin\vartheta,\quad
  \pi^{\vartheta\vartheta} = \frac{P_R}{R}\sin\vartheta
\]
and
\be
  \theta = 16\pi \int dr(P_\Lambda d\Lambda + P_R dR).
\label{Liouv-vol}
\ee
The volume terms in the super-Hamiltonian become
\bea
  \lefteqn{\frac{1}{16\pi} \int_{S_\pm}d^3x\,(NC + N^kC_k) = } \nn \\
  && \int_{S_\pm}dr\, \left\{N^r(-\Lambda P'_\Lambda + R'P_R)
    + N\left(\frac{\Lambda}{2R^2}P_\Lambda^2 - \frac{1}{R}P_\Lambda P_R -
      \frac{\Lambda R^2}{4}R^{(3)}\right)\right\},
\label{Ham-vol}
\eea
where
\be
  R^{(3)} = - 4\frac{R''}{\Lambda^2R} + 4\frac{\Lambda'R'}{\Lambda^3R} -
  2\frac{R^{\prime 2}}{\Lambda^2R^2} + \frac{2}{R^2}
\label{curv}
\ee
is the curvature scalar of the metric $q_{kl}$ and the hypersufaces $S_\pm$ are
defined by $\pm r > 0$.

The surface terms containing only the geometrical quantities are our next
task. The shell surface $\Sigma$ is defined by $r = 0$. Thus, for the normal
$m^k$ to $\Sigma \cap S$ in $S$, we have
\[
  m_k = \Lambda\delta^r_k,
\]
and the normal $\tilde{m}^\mu$ to $\Sigma$ in $M$ is
\[
  \tilde{m}_\mu = \frac{1}{\sqrt{g^{rr}}}\delta^r_\mu =
  \frac{\Lambda N}{\sqrt{N^2 - N_r^2\Lambda^{-2}}}\delta^r_\mu.
\]
Then
\[
  N^\bot = N^r\Lambda = \frac{N_r}{\Lambda},\quad \nu = \sqrt{{N^2 -
  N_r^2\Lambda^{-2}}},
\]
and
\[
  \sinh\alpha = \frac{N_r}{\Lambda\nu}.
\]

The definitions of $\lambda_{KL}$, $l_{KL}$, $\tilde{\pi}^{\bot\bot}$ and
$\tilde{\pi}^\bot_K$ yield
\[
  \lambda_{KL} = \left( \begin{array}{ll}
                 R^2, & 0 \\
                 0, & R^2\sin^2\vartheta
                 \end{array} \right),\quad \sqrt{\lambda} = R^2\sin\vartheta,
\]
\[
  l_{KL} = \frac{R'}{R\Lambda}\lambda_{KL},\quad l = \frac{2R'}{R\Lambda},
  \quad \tilde{\pi}^{\bot\bot} = \frac{2}{R^2}P_\Lambda,\quad
  \tilde{\pi}^\bot_K = 0.
\]
Hence, the surface term in the Hamiltonian (\ref{hamilt}) becomes
\be
  \int_{\Sigma\cap S}d^2\xi\,(\nu C_s + v^KC_{sK}) =
  \nu\left(\left[-P_\Lambda\sinh\alpha +
    \frac{RR'}{\Lambda}\cosh\alpha \right] + M(R) \right)_{r=0}
\label{Ham-shell}
\ee
and an analogous term in the symplectic form (\ref{varHcheck}) is
\be
  \frac{1}{16\pi}\int_{\Sigma\cap
    S}d^2\xi\,\sqrt{\lambda}\left(\frac{\dot{\lambda}}{\lambda}\delta[\alpha]
    - [\dot{\alpha}]\frac{\delta\lambda}{\lambda}\right) =
  \left(\delta[\alpha]R\dot{R} - R\delta R [\dot{\alpha}] \right)_{r=0}. 
\label{symp-shell}
\ee

The matter space $Z$ will carry the coordinates $z^1 = \Theta$, $z^2 = \Phi$,
and the mole density $h = \sin\Theta$ (in fact, any scalar factor in front of
$h$ can be swallowed by $e(n)$); the matter fields $z^A(\xi)$ will
simply be 
\[
  \Theta(t,\vartheta,\varphi) \equiv \vartheta,\quad \Phi(t,\vartheta,\varphi)
  \equiv \varphi.
\]
Thus, $z^A_K = \delta^A_K$, $\dot{z}^A = 0$, and we obtain
\[
  j^\alpha = (h,0,0),\quad p_A = 0,\quad
  n = \frac{\sin\vartheta}{\sqrt{\lambda}} = \frac{1}{R^2}.
\]
The fluid Hamiltonian is
\[
  - T^0_0 = \nu R^2\sin\vartheta\rho = \nu e\sin\vartheta,
\]
or
\be
  - \int_{\Sigma\cap S}d\vartheta d\varphi\, T^0_0 = 4\pi\nu e.
\label{Ham-matt}
\ee
We introduce the so-called {\em mass function} $M(R) := 4\pi e(R^{-2})$; the
meaning of it is the total rest mass of the shell of radius $R$ (see Ref.\
\cite{H-B}). 

As the momentum $p_A$ is identically zero, there is no contribution to the
symplectic form by the matter.

Collecting the results (\ref{Ham-vol}), (\ref{Liouv-vol}), (\ref{Ham-shell}),
(\ref{symp-shell}) and (\ref{Ham-matt}), we obtain the Hamiltonian for the
spherically symmetric system:
\bea
  {\mathcal H} & = & \int_{r<0}dr\,\left\{N^r(-\Lambda P'_\Lambda +
    R'P_R)
   + N\left(\frac{\Lambda}{2R^2}P_\Lambda^2 - \frac{1}{R}P_\Lambda
      P_R - \frac{\Lambda R^2}{4}R^{(3)}\right)\right\} \nn \\
   & + & \int_{r>0}dr\,\left\{N^r(-\Lambda P'_\Lambda +
    R'P_R)
   + N\left(\frac{\Lambda}{2R^2}P_\Lambda^2 - \frac{1}{R}P_\Lambda
      P_R - \frac{\Lambda R^2}{4}R^{(3)}\right)\right\} \nn \\
   & + & \nu\left(\left[-P_\Lambda\sinh\alpha +
     \frac{RR'}{\Lambda}\cosh\alpha\right] + M(R)\right)_{r=0} + E(\infty) +
     E(-\infty), 
\label{Ham-spher}
\eea
where $E(\pm\infty)$ is the ADM energy at $r = \pm \infty$ and $R^{(3)}$ is
given by Eq.\ (\ref{curv}). The symplectic form reads
\bea
  \Omega(\delta\mathbf{X},\dot{\mathbf{X}}) & = &
  \int_0^\infty dr\,(\delta P_\Lambda\dot{\Lambda} -
  \delta\Lambda\dot{P}_\Lambda + \delta P_R \dot{R} - \delta R\dot{P}_R) \nn
  \\
  & + & \int_{-\infty}^0 dr\,(\delta P_\Lambda\dot{\Lambda} -
  \delta\Lambda\dot{P}_\Lambda + \delta P_R \dot{R} - \delta R\dot{P}_R) \nn
  \\ 
  & + & (\delta[\alpha]R\dot{R} - R\delta R[\dot{\alpha}])_{r=0},
\label{symp-spher}
\eea
where 
\beann
  \delta\mathbf{X} & = & (\delta P_\Lambda(r), \delta\Lambda(r), \delta
  P_R(r), \delta R(r), \delta[\alpha]_{r=0}, \delta(R^2/2)_{r=0} ), \\
  \dot{\mathbf{X}} & = & (\dot{P}_\Lambda(r), \dot{\Lambda}(r), \dot{P}_R(r),
  \dot{R}(r), [\dot{\alpha}_{r=0}], (R^2/2)^{\mbox{.}}_{r=0}).
\eeann
The equation of motion can be obtained from the variation formula
\[
  \delta{\mathcal H} = \Omega(\delta\mathbf{X},\dot{\mathbf{X}}).
\]
The same equations of motion can be obtained from a Hamiltonian
action $I$, if we employ the corresponding Liouville form instead of the
symplectic one: 
\be
  I = \int dt \left( \int_{-\infty}^0 dr\,(P_\Lambda \dot{\Lambda} + P_R
    \dot{R}) +  
  \int_0^{\infty} dr\,(P_\Lambda \dot{\Lambda} + P_R \dot{R}) +
  ([\alpha]R\dot{R})_{r=0} - {\mathcal H} \right)
\label{action}
\ee 
We have assumed that the fields $N$, $N^r$, $\Lambda$, $R$, $P_\Lambda$ and
$P_R$ satisfy the usual fall-off conditions as described by Ref.\ \cite{kuch1}
in detail.

\section{Kucha\v{r}' transformation}
\label{sec:kuch}
Kucha\v{r}' transformation is a canonical transformation of the gravitational
volume variables so that the new variables can be neatly separated into the
true degrees of freedom and the variables that indicate a point in the
solution spacetime. An example is given in Ref.\ \cite{kuch1} where the
spherically symmetric gravity is studied. The transformation leads to a pair
of physical variables (one degree of freedom) and to the remaining variables
being the Schwarzschild time $T(r)$, the curvature radius $R(r)$ and the
conjugate momenta. The foliation of each spacetime solution remains completely
arbitrary. As a byproduct, the equations of motion for gravity become trivial.
This will help us to express the action (\ref{action}) through shell variables
alone without restricting the reparametrization invariance.

\subsection{Transformation zu $E$ and $P_E$}
In Ref.\ \cite{kuch1}, the transformation is performed in two steps. This
subsection goes the first one transforming the variables
$(P_\Lambda,\Lambda,P_R,R)$ to $(P_E,E,\mbox{P}_R,R)$. The transformation can 
be written as follows 
\beann
  E & = & \frac{R}{2}(1 - F_1F_2), \\
  P_E & = & \frac{\Lambda P_\Lambda}{RF_1F_2}, \\
  \mbox{P}_R & = & P_R - \frac{(F_1F_2 + 1)}{F_1F_2}\frac{\Lambda
    P_\Lambda}{2R} 
  - \frac{R}{2}\left(\ln\left|\frac{F_1}{F_2}\right|\right)', 
\eeann
where the useful abbreviations $F_1$ and $F_2$ are 
\[
  F_1 = \frac{R'}{\Lambda} + \frac{P_\Lambda}{R},\quad F_2 =
  \frac{R'}{\Lambda} - \frac{P_\Lambda}{R}.
\]
The inverse transformation is
\bea
  \Lambda & = & \sqrt{-FP_E^2 + F^{-1}R^{\prime 2}}, \nn \\
  P_\Lambda & = & \frac{RFP_E}{\Lambda},
\label{Plambda} \\
  P_R & = & \mbox{P}_R + \frac{F+1}{2}P_E +
  \frac{R}{F}\left(\frac{FP_E}{R'}\right)'\frac{R^{\prime 2}}{\Lambda^2}, \nn
\eea
where
\be
  F := \frac{R - 2E}{R},
\label{F}
\ee
and $\Lambda$ on the R. H. S.'s must be expressed with the help of the first
equation. 

The following important relations hold \cite{kuch1}
\be
  F = F_1F_2,\quad P_E = - T'.
\label{Tprime}
\ee
The transformation of the volume part of the Liouville form has the form
\cite{kuch1} 
\be
  P_\Lambda d\Lambda + P_RdR = P_EdE + \mbox{P}_RdR +
  \left(\frac{RdR}{2}\ln\left|\frac{F_1}{F_2}\right|\right)' + \ldots,
\label{kuch-liouv}
\ee
where the dots denote a differential of some function on the phase space,
which can be discarded. In Ref.\ \cite{kuch1}, the $r$-derivative term on the
R. H. S. of Eq.\ (\ref{kuch-liouv}) could also be thrown away because the
asymptotic values of the differentiated function vanished. In our case,
however, this term gives a non-trivial contribution to the shell part of the
Liouville form: 
\bea
  \lefteqn{\int_{-\infty}^0dr(R_\Lambda d\Lambda + P_RdR) +
  \int_0^{\infty}dr(P_\Lambda d\Lambda + P_RdR) = } \nn \\
  && \int_{-\infty}^0dr(R_E dE + \mbox{P}_RdR) +
  \int_0^{\infty}dr(R_E dE + \mbox{P}_RdR) + \nn \\
  && \left(\left[\ln\sqrt{\left|\frac{F_2}{F_1}\right|}\right]
    RdR\right)_{r=0}.   
\label{log}
\eea 
Let us study the geometrical meaning of the last term. The meaning of any
quantity in the canonical formalism is given by the role it plays in the
classical solutions. We can, therefore, assume that the canonical equations are
satisfied. The only canonical equation we need is
\[
  P_\Lambda = - \frac{R}{N}(\dot{R} - R'N^r);
\]
it implies that
\be
  F_{1,2} = \left(\frac{1}{\Lambda}\frac{\partial}{\partial r} \mp
    \frac{1}{N}\left(\frac{\partial}{\partial t} - N^r\frac{\partial}{\partial
        r} \right)\right)R.
\label{F1F2}
\ee
We also have
\beann
  \frac{\partial}{\partial t} & = & N {n} + N^r\Lambda{m}, \\
  \frac{\partial}{\partial r} & = & \Lambda {m},
\eeann
where ${m}$ is the right-oriented unit vector normal to $S\cap\Sigma$ and
tangential to $S$, and ${n}$ is the future-oriented unit vector normal to
$S$ at $S\cap\Sigma$ (these vectors carry, of course, the indices $\pm$ that we
are leaving out provisionally); we call $({n},{m})$ {\em foliation
  frame}. It follows that
\[
  F_{1,2} = (m^\mu \mp n^\mu)\frac{\partial R}{\partial x^\mu}.
\]
Clearly, $m^\mu \mp n^\mu$ are radial null vectors; $F_1$ vanishes at the
left-going (past) and $F_2$ at the right-going (future) horizon (cf.\ Ref.\ 
\cite{kuch1}). The meaning of the logarithm in Eq. (\ref{log}) will be evident
if we introduce the so-called {\em Schwarzschild frame} $(n_S,m_S)$ defined by
the conditions: the frame $(n_S,m_S)$ is orthonormal, future- and
right-oriented, and such that at least one of its vectors (as a differential
operator) annihilates the function $R$. The horizons divide the Kruskal
manifold into four quadrants $Q_I$--$Q_{IV}$. We identify them as follows:
$Q_I$ is adjacent to the right infinity, $Q_{II}$ to the left one, $Q_{III}$
to the future singularity and $Q_{IV}$ to the past one. The Schwarzschild
frame is well-defined only {\em inside} the four quadrants, and its components
with respect to the Schwarzschild coordinates $T$ and $R$ there are given by
the Table \ref{tab:schw}.
\begin{table}
\centering
\begin{tabular}{l||l|l|} 
   & $n_S$ & $m_S$ \\[.1cm] \hline\hline
$Q_I$ & $(\frac{1}{\sqrt{|F|}},0)$ & $(0,\sqrt{|F|})$ \\ \hline
$Q_{II}$ & $(-\frac{1}{\sqrt{|F|}},0)$ & $(0,-\sqrt{|F|})$ \\ \hline
$Q_{III}$ & $(0,-\sqrt{|F|})$ & $(\frac{1}{\sqrt{|F|}},0)$ \\ \hline 
$Q_{IV}$ & $(0,\sqrt{|F|})$ & $(-\frac{1}{\sqrt{|F|}},0)$ \\ \hline
\end{tabular}
\caption{Components of the Schwarzschild frame}
\label{tab:schw}
\end{table}
Let us define the angle $\beta$ as the hyperbolic rotation angle from the
Schwarzschild to the foliation frame:
\beann
  n & = & n_S\cosh\beta + m_S\sinh\beta, \\
  m & = & n_S\sinh\beta + m_S\cosh\beta.
\eeann
Then,
\[
  m \mp n = \mbox{e}^{\mp\beta}(m_S \mp n_S),
\]
and 
\[
  (m^\mu \mp n^\mu)\partial_\mu R = \mbox{e}^{\mp\beta}(m^\mu_S \mp
  n^\mu_S)\partial_\mu R. 
\]
Working with the Table \ref{tab:schw}, we obtain from it that
\[
  \left|\frac{F_2}{F_1}\right| = \mbox{e}^{2\beta}
\]
in all quadrants. The final result is, therefore, simply
\be
  \left[\ln\sqrt{\left|\frac{F_2}{F_1}\right|}\right]RdR = [\beta]RdR,
\label{beta}
\ee
and the first step of Kucha\v{r}' transformation changes the shell part of
the Liouville form as follows 
\[
  ([\alpha]RdR)_{r=0} \rightarrow ([\alpha + \beta]RdR)_{r=0}.
\]
The definition (\ref{alpha}) implies that $\alpha$ is the angle of the
hyperbolic rotation from the foliation frame to the {\em shell frame}
$(\tilde{n},\tilde{m})$. Here, the vector $\tilde{n}$ is future-oriented,
orthogonal to $S\cap\Sigma$, and tangential to $\Sigma$, $\tilde{m}$ is
right-oriented and orthogonal to $\Sigma$ at $S\cap\Sigma$. We have from Eq.\ 
(\ref{alpha}):
\beann
  \tilde{n} & = & n\cosh\alpha + m\sinh\alpha, \\
  \tilde{m} & = & n\sinh\alpha + m\cosh\alpha.
\eeann
Thus, $\alpha + \beta$ is the angle of the hyperbolic
rotation from the Schwarzschild  to the shell frame. Let us define:
\[
  P = (\alpha + \beta)R;
\]
$P$ is independent of the foliation and {\em singular} at the horizons.

The constraints $C$ and $C_r$ are written down in terms of the new variables
in Ref.\ \cite{kuch1}. More interesting for us is that these constraints can be
replaced by an equivalent pair $C_1$ and $C_2$ that is much simpler
\cite{kuch1}: 
\[
  C_1 = E'(r),\quad C_2 = \mbox{P}_R(r).
\]

The shell constraint contains the expression 
\[
  \tilde{C} := -P_\Lambda\sinh\alpha + \frac{RR'}{\Lambda}\cosh\alpha
\]
that reads in the new variables as follows (cf.\ Eqs.\
(\ref{Plambda}) and (\ref{Tprime})):
\[
  \tilde{C} = \frac{RFT'}{\Lambda}\sinh\alpha + \frac{RR'}{\Lambda}\cosh\alpha.
\]
This can be expressed by means of the angle $\alpha + \beta = P/R$. The
foliation frame has the following components with respect to the Schwarzschild
coordinates:
\[
  n = \left(\frac{R'}{F\Lambda},\frac{FT'}{\Lambda}\right),\quad m =
  \left(\frac{T'}{\Lambda}, \frac{R'}{\Lambda}\right),
\]
and this holds in all quadrants. It follows that $\sinh\beta$ and $\cosh\beta$
is related to $T'$ and $R'$ as given in Table \ref{tab:beta}.
\begin{table}
\centering
\begin{tabular}{l||l|l|l|l|}
             & $Q_I$ & $Q_{II}$ & $Q_{III}$ & $Q_{IV}$ \\[.2cm] \hline\hline
$\sinh\beta$ & $\frac{T'\sqrt{|F|}}{\Lambda}$ &
$-\frac{T'\sqrt{|F|}}{\Lambda}$ & $- \frac{R'}{\Lambda\sqrt{|F|}}$ &
$\frac{R'}{\Lambda\sqrt{|F|}}$ \\[.2cm] \hline
$\cosh\beta$ & $\frac{R'}{\Lambda\sqrt{|F|}}$ &
$-\frac{R'}{\Lambda\sqrt{|F|}}$ 
& $\frac{T'\sqrt{|F|}}{\Lambda}$ & $-\frac{T'\sqrt{|F|}}{\Lambda}$ \\[.2cm]
\hline 
\end{tabular}
\caption{$\sinh\beta$ and $\cosh\beta$ by means of the canonical variables}
\label{tab:beta}
\end{table}
The following notation will enable us to write formulas valid in all quadrants
simultaneously: let
\[
  \mbox{sh}_+x := \cosh x,\quad \mbox{sh}_-x := \sinh x,
\]
and let $a$ and $b$ be signs defined by Table \ref{tab:ab}.
\begin{table}
\centering
\begin{tabular}{l||l|l|l|l|}
             & $Q_I$ & $Q_{II}$ & $Q_{III}$ & $Q_{IV}$ \\ \hline\hline
$a$ & $+$ & $+$ & $-$ & $-$ \\ \hline
$b$ & $+$ & $-$ & $-$ & $+$ \\ \hline
\end{tabular}
\caption{The signs $a$ and $b$}
\label{tab:ab}
\end{table}
Then,
\[
 \tilde{C} = bR\sqrt{|F|}\ \mbox{sh}_a\frac{P}{R}.
\]

To summarize: the Hamiltonian action $I$ of the system reads:
\bea
  I & = & \int dr \left\{ \int_{-\infty}^0 dr (P_E\dot{E} + \mbox{P}_R\dot{R}
    - N_1C_1 - N_2C_2)\right. \nn \\
  & + & \int_0^{\infty} dr (P_E\dot{E} + \mbox{P}_R\dot{R} - N_1C_1 -
  N_2C_2) + \left([P]\dot{R}\right)_{r=0} \nn \\
  & + & \left.  \nu\left(\left[
  bR\sqrt{|F|}\ \mbox{sh}_a\frac{P}{R}\right] + M(R) \right)_{r=0}
  - E(\infty) - E(-\infty)\right\},
\label{ab-action}
\eea
where $E(\infty)$ and $E(-\infty)$ are the ADM masses at each of the spacelike
infinities.

\subsection{Transformation to $T$ and $P_T$}
The second step of Kucha\v{r}' transformation concerns the variables $E$ and
$P_E$ and the boundary terms at the infinities. We shall use a slightly
modified version of Kucha\v{r}' procedure in this section.

Each given boundary term at the infinities assume some particular boundary
condition; in our case, the lapse function $N(\pm\infty)$ must be kept equal
to 1. We need more freedom, however. Such a freedom is achieved in Ref.\ 
\cite{kuch1} by parametrizing the system at the infinities. This can be done
by introducing the coordinates $T(\pm\infty)$ of the hypersurface $S_t$ at $r =
\pm\infty$. In Ref.\ \cite{kuch1}, it is shown that
\[
  N(\pm\infty) = \pm\dot{T}(\pm\infty)
\]
and the term $E(\infty) + E(-\infty)$ in the Hamiltonian (\ref{Ham-spher}) or
in the action (\ref{ab-action}) is to be replaced by $E(\infty)\dot{T}(\infty)
- E(-\infty)\dot{T}(-\infty)$. Then, all variations can be performed,
including arbitrary variation of $N$ at both infinities, and the result are
valid equations \cite{kuch1}.

The term $E(\infty)\dot{T}(\infty) - E(-\infty)\dot{T}(-\infty)$ in the action
can of course be considered as a part of the Liouville form; thus, the
parametrized action contains the Liouville term:
\bea
  \dot{\theta} & = &\int_{-\infty}^0dr(P_E\dot{E} + \mbox{P}_R\dot{R}) +
  \int_0^{\infty}dr(P_E\dot{E} + \mbox{P}_R\dot{R}) \nn \\ 
  & + & ([\psi]R\dot{R})_{r=0} - E(\infty)\dot{T}(\infty) +
  E(-\infty)\dot{T}(-\infty).
\label{theta}
\eea

The next step is to introduce the new variable $T(r)$ that satisfies the
relation $P_E = -T'$ (see Eq.\ (\ref{Tprime})) and to find the corresponding
conjugate momentum. This can be done by a transformation that
concerns only the variables $E$, $P_E$, $E(\pm\infty)$ and $T(\pm\infty)$. The
relevant parts of the Liouville form are
\[
  \dot{\theta}_+ = \int_0^{\infty}dr\,P_E\dot{E} - E(\infty)\dot{T}(\infty)
\]
and
\[
  \dot{\theta}_- = \int_{-\infty}^0dr\,P_E\dot{E} + E(-\infty)\dot{T}(-\infty).
\]
Let us substitute $-T'$ for $P_E$ in $\dot{\theta}_+$ and transfer the primes
and dots as follows
\[
  \dot{\theta}_+ = - \int_0^{\infty}dr\,E'\dot{T} - (E\dot{T})_{r=0} +
  \left(-\int_0^{\infty}dr\,ET' \right)^{\mbox{.}}.
\]
Similarly one obtains for $\dot{\theta}_-$:
\[
   \dot{\theta}_- = - \int_{-\infty}^0dr\,E'\dot{T} + (E\dot{T})_{r=0} +
   \left(-\int_{-\infty}^0dr\,ET' \right)^{\mbox{.}}.
\]
Hence, $P_T = - E'$, and the constraints simplify even further:
\be
  C_1 = - P_T,\quad C_2 = \mbox{P}_R.
\label{new-constr}
\ee
If we introduce the notation
\[
  \lim_{r=0\pm}T(r) = T_\pm,\quad \lim_{r=0\pm}E(r) = E_\pm,
\]
then the action (\ref{ab-action}) in the new variables reads:
\bea
  I & = & \int dt\left\{\int_{-\infty}^0dr(P_T\dot{T} + \mbox{P}_R\dot{R}) +
    \int_0^{\infty}dr(P_T\dot{T} + \mbox{P}_R\dot{R})\right. \nn \\
    & + & ([P]\dot{R} - E_+\dot{T}_+ + E_-\dot{T}_-)_{r=0} \nn \\
    & - & \int_{-\infty}^0dr(N_1C_1 + N_2C_2) - \int_0^{\infty}dr(N_1C_1 +
    N_2C_2) \nn \\
    & - & \left.\nu\left(\left[bR\sqrt{|F|}\
    \mbox{sh}_a\left(\frac{P}{R}\right)\right] + M(R)\right)_{r=0}\right\}, 
\label{action2}
\eea
where $C_1$ and $C_2$ are given by Eq.\ (\ref{new-constr}) and $a$ and $b$ by
Table \ref{tab:ab}.

The dynamical equations for the variables $T$, $R$, $P_T$ and $\mbox{P}_R$
describing the gravitational field around the shell that result from the
action (\ref{action2}) are: 
\be
  \dot{T} = -N_1,\quad \dot{R} = N_2,
\label{dot-N}
\ee and
\be
  P_T = 0,\quad \mbox{P}_R = 0.
\label{constr-fin}
\ee
The first pair (\ref{dot-N}) does not impose any limitations on $\dot{T}$ and
$\dot{R}$ because the Lagrange multipliers $N_1$ and $N_2$ are arbitrary. The
second pair (\ref{constr-fin}) implies that $E(r)$ is constant along each
slice, $E(t,r) = E_+(t)$ and $E(t,r) = E_-(t)$. This together with $\mbox{P}_R
= 0$ does not even imply that the spacetime outside the shell is Schwarzschild
one.

The non-trivial part of the dynamics is completely contained in the shell
equations. The shell Hamiltonian depends on the variables $\nu$ $R$, $E_\pm$,
$P_\pm$ and on the discrete variables $a_\pm$ and $b_\pm$. It does not
depend on $T_\pm$! It follows immediately that $\dot{E}_\pm = 0$. This,
together with the volume equations (\ref{dot-N}) and (\ref{constr-fin}) is
equivalent to Schwarzschild solution being the spacetime outside the shell.

We can, therefore, replace the action (\ref{action2}) by an effective shell
action of the form:
\be
  I_s = \int dt\left(P_+\dot{R} - E_+\dot{T}_+ - P_-\dot{R} +
    E_-\dot{T}_- - \nu C_s \right),
\label{shell-act}
\ee
where
\be
  C_s = b_+R\sqrt{|F_+|}\ \mbox{sh}_{a_+}\frac{P_+}{R}
    - b_-R\sqrt{|F_-|}\ \mbox{sh}_{a_-}\frac{P_-}{R} +
      M(R),
\label{shell-H}
\ee
is the super-Hamiltonian of the shell and $F_\pm$ are given by Eq.\ (\ref{F}).
We interpret the solutions $E_\pm$, $T_\pm(t)$, $P_\pm(t)$ and $R(t)$ as
embedding formulas in two Schwarzschild spacetimes with energies $E_\pm$ and
coordinates $(T,R)$.

The discrete variables $a_\pm$ and $b_\pm$ describe the different sectors of
the extended phase space. If the shell crosses a horizon in the spacetime to
its left or right, some of the signs will change. There are 16 sectors; some
of these, however, will have empty intersection with the constraint surface.
Observe that $a_\pm$ is not an independent variable, but a function of $R$ and
$E_\pm$: 
\be 
  a_\pm = \mbox{sign}F_\pm.
\label{a}
\ee
The action (\ref{shell-act}) describes the motion inside the sectors and it
becomes singular at sector boundaries. The variables $P_\pm$ and $T\pm$
diverge and $a_\pm$ and $b_\pm$ are not defined at the boundaries.

\section{Properties of the shell action}
\label{sec:action}
In this section, we study the properties of the action
(\ref{shell-act}). We derive the equations of motion, clarify the structure of
constraints and reveal a geometrical meaning of the super-Hamiltonian.

\subsection{The equations of motion}
Let us vary the action (\ref{shell-act}). The variation of $\nu$ gives the
Hamiltonian constraint:
\be
  b_+R\sqrt{|F_+|}\ \mbox{sh}_{a_+}\frac{P_+}{R}
  - b_-R\sqrt{|F_-|}\ \mbox{sh}_{a_-}\frac{P_-}{R} + M(R) = 0,
\label{Hconstr}
\ee
and the variations of $P_\pm$ result in
\bea
  \frac{\dot{R}}{\nu} & = & b_+\sqrt{|F_+|}\
    \mbox{sh}_{-a_+}\frac{P_+}{R},
\label{Rdot1} \\
  \frac{\dot{R}}{\nu} & = & b_-\sqrt{|F_-|}\
  \mbox{sh}_{-a_-}\frac{P_-}{R},
\label{Rdot2}
\eea
which implies another constraint:
\be
  b_+\sqrt{|F_+|}\ \mbox{sh}_{-a_+}\frac{P_+}{R}
  - b_-\sqrt{|F_-|}\ \mbox{sh}_{-a_-}\frac{P_-}{R} = 0.
\label{chi}
\ee
The variation with respect to $E_\pm$ yields:
\[
  - \dot{T}_\pm - \nu
  b_\pm\frac{R}{2\sqrt{|F_\pm|}}\ \mbox{sh}_{a_\pm}\frac{P_\pm}{R}\ 
  \frac{\partial|F_\pm|}{\partial E_\pm}.
\]
For the calculation of the derivative of $|F_\pm|$, we take Eq.\ (\ref{a})
into account:
\[
  \frac{\partial|F_\pm|}{\partial E_\pm} = - \frac{2a_\pm}{R}.
\]
Then, the following equation results:
\be
  \frac{\dot{T}_\pm}{\nu} = \frac{a_\pm b_\pm}{\sqrt{|F_\pm|}}\ 
  \mbox{sh}_{a_\pm}\frac{P_\pm}{R}.
\label{Tdot}
\ee
The variation of $T_\pm$ leads to
\be
  \dot{E}_\pm = 0.
\label{Edot}
\ee
Finally, varying $R$, we obtain
\be
  -\dot{P}_+ + \dot{P}_- - \nu \frac{\partial C_s}{\partial R}.
\label{Pdot}
\ee
Eqs.\ (\ref{Hconstr})--(\ref{Pdot}) form the complete set of
dynamical equations for the shell. Some discussion of these equations is in
order. 

First, we show that Eq.\ (\ref{Pdot}) is a consequence of
Eqs.\ (\ref{Hconstr})--(\ref{Rdot2}), (\ref{Edot}) and of $\dot{R} \neq
0$ (the last relation is generically satisfied along each trajectory). Indeed,
the time derivative of the super-Hamiltonian $C_s$ must vanish as 
a consequence of Eq.\ (\ref{Hconstr}):
\[
  \frac{\partial C_s}{\partial R}\dot{R} + \frac{\partial C_s}{\partial
  E_+}\dot{E}_+ + \frac{\partial C_s}{\partial E_-}\dot{E}_- + \frac{\partial
  C_s}{\partial P_+}\dot{P}_+ + \frac{\partial C_s}{\partial P_-}\dot{P}_- =
  0. 
\]
The second and third terms on the L. H. S. vanish because of
Eq. (\ref{Edot}). For the last two terms, we obtain from Eqs. (\ref{Rdot1})
and (\ref{Rdot2}): 
\[
 \frac{\partial C_s}{\partial P_+} = \frac{\dot{R}}{\nu},\quad \frac{\partial
 C_s}{\partial P_-} = - \frac{\dot{R}}{\nu}.
\]
Hence,
\[
  \left(\frac{\partial C_s}{\partial R} + \frac{\dot{P_+}}{\nu} -
  \frac{\dot{P_-}}{\nu}\right)\dot{R} = 0,
\]
and this shows the claim.

Second, manipulating Eqs.\ (\ref{Rdot1}), (\ref{Rdot2}) and (\ref{Tdot}), we
arrive at 
\be
  -|F_\pm|\left(\frac{\dot{T}_\pm}{\nu}\right)^2 +
  \frac{1}{|F_\pm|}\left(\frac{\dot{R}}{\nu}\right)^2 = -\ 
  \mbox{sh}^2_{a_\pm}\frac{P_\pm}{R} +
  \mbox{sh}^2_{-a_\pm}\frac{P_\pm}{R}.
\label{met}
\ee
A useful identity is
\be
  \mbox{sh}_a(x+y) = \cosh x\ \mbox{sh}_ay + \sinh x\ \mbox{sh}_{-a}y,
\label{sh-id}
\ee
which can easily be derived from the definition,
\be
  \mbox{sh}_ax = \frac{\mbox{e}^x + a\mbox{e}^{-x}}{2},
\label{sh-def}
\ee
and which implies that $\mbox{sh}^2_ax - \mbox{sh}^2_{-a}x = a$, independently
of $x$. Thus, the R. H. S. of Eq.\ (\ref{met}) is $-a_\pm$. Multiplying the
equation by $a_\pm$ and using Eq.\ (\ref{a}) yield
\be
  - F_\pm\left(\frac{\dot{T}_\pm}{\nu}\right)^2 +
  \frac{1}{F_\pm}\left(\frac{\dot{R}}{\nu}\right)^2 = -1.
\label{Tdot2}
\ee
This is the `time equation' (see Ref.\ \cite{H-B}) saying that
\[
  \left(\left(\frac{\dot{T}_\pm}{\nu}\right),
  \left(\frac{\dot{R}}{\nu}\right)\right) 
\]
is a unit timelike vector. It also implies  that
\be
  F_\pm\frac{\dot{T}_\pm}{\nu} = \tau_\pm\sqrt{F_\pm +
    \left(\frac{\dot{R}}{\nu}\right)^2}, 
\label{e}
\ee
where 
\be
  \tau_\pm := \mbox{sign}\left(F_\pm\frac{\dot{T}_\pm}{\nu}\right).
\label{eta}
\ee

Finally, we obtain from Eqs.\ (\ref{Hconstr}) and (\ref{Tdot}):
\[
  F_+\frac{\dot{T}_+}{\nu} - F_-\frac{\dot{T}_-}{\nu} = - \frac{M(R)}{R};
\]
substituting into this equation from (\ref{e}) yields the `radial equation':
\be
  -\tau_+\sqrt{F_+ + \left(\frac{\dot{R}}{\nu}\right)^2} + \tau_-\sqrt{F_- +
    \left(\frac{\dot{R}}{\nu}\right)^2} = \frac{M(R)}{R}.
\label{israel}
\ee 
This is Israel's equation for spherically symmetric shells written in a
way that is valid for all sectors in the case of future-oriented shell motion
(see Ref.\ \cite{H-B}). Thus, the dynamical equations implied by the action
(\ref{shell-act}) are as they should be.

\subsection{Structure of the constraints}
Two constraint functions have been obtained directly from the action
(\ref{shell-act}) by varying it: the super-Hamiltonian $C_s$ and the L. H. S.
of Eq.\ (\ref{chi}), which we denote by $\chi$. The Lagrange multiplier that
gives $C_s$ is $\nu$, that for $\chi$ is $\bar{P}$, defined by
\[
  \bar{P} := \frac{P_+ + P_-}{2}.
\]

The Poisson bracket between $C_s$ and $\chi$ requires a longer calculation; we
quote just the result:
\beann
 \lefteqn{\{\chi,C_s\} \approx} \\ 
 && \mbox{} - \frac{E_+ - E_-}{2R^2}\left(1 +
 \frac{a_+b_+b_-}{\sqrt{|F_+F_-|}}\ \mbox{sh}_{a_+a_-}\frac{[P]}{R}
 - \frac{2\mbox{sign}B}{\sqrt{B^2 - A^2}}(M' - M/R)\right) \\ && \mbox{} -
 \frac{a_-b_+b_-\mbox{sign}B}{R^2\sqrt{B^2 -
 A^2}}\ \mbox{arctanh}\frac{A}{B}\ M(R)\sqrt{|F_+F_-|}\
 \mbox{sh}_{-a_+a_-}\frac{[P]}{R}, 
\eeann
where
\bea
  A & := & b_+\sqrt{|F_+|}\ \mbox{sh}_{a_+}\frac{[P]}{2R} -
  a_-b_-\sqrt{|F_-|}\ \mbox{sh}_{a_-}\frac{[P]}{2R},
\label{X} \\
  B & := & b_+\sqrt{|F_+|}\ \mbox{sh}_{-a_+}\frac{[P]}{2R} -
  a_-b_-\sqrt{|F_-|}\ \mbox{sh}_{-a_-}\frac{[P]}{2R},
\label{Y}
\eea
and $[P] = P_+ - P_-$ is the momentum conjugate to $R$. We have used the
constraints in calculating the bracket, so the equality is only weak
(`$\approx$'). The Poisson bracket is non-zero at the constraint surface, so
our system cannot be purely first-class and the value of some Lagrange
multipliers will be determined by the equations of motion (see, eg.\
\cite{H-T}). Clearly, it is $\bar{P}$ which is determined, for $\chi$ depends
on it and can, therefore be used to calculate it:
\[
  \chi = A\cosh\frac{\bar{P}}{R} + B\sinh\frac{\bar{P}}{R},
\]
or
\be
  \bar{P} = - R\ \mbox{arctanh}\frac{A}{B}.
\label{psi}
\ee

The Lagrange multiplier $\nu$ is not restricted by the equations of
motion. This means that the system is mixed, containing both first- and
second-class constraints. To prove that, we extend the phase space by another
conjugate pair, $(\bar{P}, \pi)$ and constraint the momentum $\pi$ to be zero,
\[
  \pi = 0.
\]
This constraint must be enforced by another Lagrange multiplier, $\tilde{\nu}$,
say, and the corresponding additional term in the action is
$-\tilde{\nu}\pi$. The new system is clearly equivalent to the old one, but it
has three constraints, $C_s$, $\chi$ and $\pi$. We obtain easily:
\[
  \{\pi,C_s\} = \frac{\partial C_s}{\partial\bar{P}} = \chi,
\]
 and 
\[
  \{\chi,\pi\} = \frac{\partial \chi}{\partial\bar{P}} = - \frac{M(R)}{R^2} +
  \frac{1}{R^2}C_s. 
\]
Thus, the pair $(\chi,\pi)$ represents the second-class part of the constraint
system, and a modification $\tilde{C}_s$ of $C_2$ defined by
\[
  \tilde{C}_s := C_s + \frac{\{\chi,C_s\}R^2}{M(R)}\,\pi
\]
has weakly vanishing Poisson brackets with both $\chi$ and $\pi$. The
equations $C_s = 0$ and $\pi = 0$ are primary constraints and $\chi = 0$ 
is a secondary constraint.

Let us write down the action of the extended system:
\be
  I_s^{e} = \int dt\left( [P]\dot{R} - E_+\dot{T}_+ + 
    E_-\dot{T}_- + \pi\dot{\bar{P}} - \tilde{\nu}\pi - \nu C_s \right), 
\label{shell-acte}
\ee
where we have to substitute $\bar{P} \pm [P]/2$ for $P_\pm$. The method of
Dirac's brackets can be applied to $I_s^e$. An (equivalent) alternative is to
get rid of $\bar{P}$ by solving the constraint $\chi = 0$ for it and inserting
the solution back into the action (\ref{shell-act}).

\subsection{Partial reduction}
In this subsection, we reduce the system partially by substituting Eq.\ 
(\ref{psi}) for $\bar{P}$ into the action (\ref{shell-act}). 

First, we make the dependence of $C_s$ on $\bar{P}$ explicit:
\[
  C_s = RA\sinh\frac{\bar{P}}{R} + RB\cosh\frac{\bar{P}}{R} + M(R),
\]
where $A$ and $B$ are given by Eqs.\ (\ref{X}) and (\ref{Y}). Eq.\ (\ref{psi})
can be written in the form
\beann
  \sinh\frac{\bar{P}}{R} & = & - \mbox{sign}B\ \frac{A}{\sqrt{B^2 - A^2}}, \\
  \cosh\frac{\bar{P}}{R} & = & \mbox{sign}B\ \frac{B}{\sqrt{B^2 - A^2}}.
\eeann
Hence, we obtain for $C_s$
\[
  C_s = \mbox{sign}B\,R\sqrt{B^2 - A^2} + M(R).
\]
Clearly, the constraint surface intersects only those sectors, where the
following conditions are satisfied:
\be
  \mbox{sign}B = -1,\quad |B| > |A|.
\label{condY}
\ee
The definitions (\ref{X}) and (\ref{Y}) lead to
\[
  B^2 - A^2 = F_+ + F_- -
  2a_-b_+b_-\sqrt{|F_+F_-|}\ \mbox{sh}_{a_+a_-}\frac{[P]}{R}, 
\]
where we have used the identity (\ref{sh-id}); we obtain the partially
reduced su\-per-Ha\-mil\-ton\-ian, which we denote by $C^r_s$:
\be
  C^r_s = R\ \mbox{sign}B\sqrt{F_+ + F_- -
  2a_-b_+b_-\sqrt{|F_+F_-|}\ \mbox{sh}_{a_+a_-}\frac{[P]}{R}} + M(R);
\label{red-cs}
\ee
the corresponding action, which is independent of $\bar{P}$ and implies only
one constraint, reads:
\be
  I_s^r  =  \int dt\left\{[P]\dot{R} - E_+\dot{T}_+ +
    E_-\dot{T}_- - \nu C^r_s \right\}. 
\label{shell-actr}
\ee
The super-Hamiltonian (\ref{red-cs}) in the four sectors where $a_+ = a_- = 1$
seems to be the same as the zero cosmological constant case of the
super-Hamiltonian derived in Ref.\ \cite{A-A-B-S}.

The expression under the square root in Eq.\ (\ref{red-cs}) reminds of the
Cosine Theorem, and, indeed, it has a simple geometrical
interpretation. Consider the vector $\xi$ generating the Schwarzschild time
shift. There is a simple expression for $\xi$ in terms of the Schwarzschild
frame, because one leg of this frame is always parallel to $\xi$; for each
quadrant, $\xi$ is given by Table \ref{tab:xi}.
\begin{table}
\centering
\begin{tabular}{l||l|l|l|l|}
  & $Q_I$ & $Q_{II}$ & $Q_{III}$ & $Q_{IV}$ \\[.2cm] \hline\hline
$\xi$ & $n_S\sqrt{|F|}$ & $-n_S\sqrt{|F|}$ & $m_S\sqrt{|F|}$ &
$-m_S\sqrt{|F|}$ \\[.2cm] \hline
\end{tabular}
\caption{The vector $\xi$ and the Schwarzschild frame}
\label{tab:xi}
\end{table}
Let us find the components of $\xi$ with respect to the shell frame using the
transformation between the shell and the Schwarzschild frame:
\beann
  n_S & = & \tilde{n}\cosh\frac{P}{R} - \tilde{m}\sinh\frac{P}{R}, \\
  m_S & = & -\tilde{n}\sinh\frac{P}{R} + \tilde{m}\cosh\frac{P}{R}.
\eeann
The result can be summarized by the formula
\be
  \xi = \tilde{n}b\sqrt{|F|}\ \mbox{sh}_a\frac{P}{R} -
  \tilde{m}b\sqrt{|F|}\ \mbox{sh}_{-a}\frac{P}{R},
\label{xi-tilde}
\ee
which is valid in all quadrants; we have left out the indices $\pm$. Comparing
Eq.\ (\ref{xi-tilde}) with the original form of the constraints $C_s$ and
$\chi$, we can see immediately that
\[
  C_s = R(\xi^0_{(+)} - \xi^0_{(-)}) + M(R)
\]
and
\[
  \chi = \xi^1_{(+)} - \xi^1_{(-)},
\]
where the shell frame components $\xi^0$ and $\xi^1$ of the vector $\xi$ are
given by Eq.\ (\ref{xi-tilde}). The geometrical meaning of the constraint
$\chi = 0$ is, therefore, that the space component of the `vector difference'
$\xi_+ - \xi_-$ vanishes, and of $C_s = 0$ that the time component of this
`vector difference' equals $-M(R)/R$.

In the case that $\chi = 0$, we have
\[
  |\xi_+ - \xi_-| = |\xi^0_{(+)} - \xi^0_{(-)}|,
\]
where $|\xi_+ - \xi_-|$ is the `length' of the `vector' $\xi_+ - \xi_-$,
defined by
\[ 
  |\xi_+ - \xi_-| = \sqrt{|-(\xi^0_{(+)} - \xi^0_{(-)})^2 +
  (\xi^1_{(+)} - \xi^1_{(-)})^2|}. 
\]
It follows that
\[
  C_s = R\ \mbox{sign}(\xi^0_{(+)} - \xi^0_{(-)})|\xi_+ - \xi_-| + M(R).
\]

Let us calculate the value of $(\xi_+ - \xi_-)^2$ using Eq.\ (\ref{xi-tilde}).
The result is:
\[
 -(\xi^0_{(+)} - \xi^0_{(-)})^2 +
  (\xi^1_{(+)} - \xi^1_{(-)})^2 = -F_+ - F_- +
  2a_-b_+b_-\sqrt{|F_+F_-|}\ \mbox{sh}_{a_+a_-}\frac{[P]}{R}. 
\]
This coincides, up to the sign, with the expression under the square root in
Eq.\ (\ref{red-cs}). It is also clear that the constraint $\chi = 0$ must
imply, first, that
\[
  \mbox{sign}B = \mbox{sign}(\xi^0_{(+)} - \xi^0_{(-)})
\]
and, second, that $B^2 - A^2 > 0$, if the vector difference $\xi_+ - \xi_-$ is
timelike. This finishes the clarification of a geometrical meaning of the
constraints.

\section{Matching the sectors}
The actions (\ref{shell-act}), (\ref{shell-acte}) and (\ref{shell-actr}) are
singular at each horizon $R = 2E_\pm$, because the coordinate $T_\pm$ and the
momentum $P_\pm$ diverge. Thus, the actions can be used only inside the 16
sectors; they do not say, at least directly, what happens at the boundary.

The form of the singularity in $P_\pm$ can be inferred from Eq.\
(\ref{xi-tilde}): both the vector $\xi$ and the shell frame
$(\tilde{n},\tilde{m})$ are smooth objects, so the components are to be smooth,
too. It follows that 
\be
  \mbox{sh}_{a_\pm}\frac{P_\pm}{R} \sim \frac{1}{\sqrt{|F_\pm|}},\quad 
  \mbox{sh}_{-a_\pm}\frac{P_\pm}{R} \sim \frac{1}{\sqrt{|F_\pm|}}
\label{singP}
\ee
at the horizons.

This section will be based on a transformation of the extended action
(\ref{shell-acte}) that may be interesting for other purposes, too. First, we
introduce the variables $R_\pm$ by
\bea
  R & = & \frac{R_+ + R_-}{2},
\label{Rbar} \\
  \pi & = & - R_+ + R_-.
\label{[R]}
\eea
The meaning of the variables $R_+$ and $R_-$ is simply that  they give the
values of the function $R$ at the shell from the right and from the left,
respectively. Thus, the constraint $\pi = 0$ is nothing but the only remaining
continuity condition from (\ref{gamma-g}). Let us substitute Eqs.\
(\ref{Rbar}) and (\ref{[R]}) into the Liouville part of the action $I^e_s$:
\beann
  \lefteqn{[P]\dot{\bar{R}} - [R]\dot{\bar{P}} - E_+\dot{T}_+ + E_-\dot{T}_-
    =} \\ 
  && [P]\dot{\bar{R}} + \bar{P}[\dot{R}] - E_+\dot{T}_+ + E_-\dot{T}_- -
  (\bar{P}[R])^{\mbox{.}} = \\
  && P_+\dot{R}_+ - E_+\dot{T}_+ - P_-\dot{R}_- + E_-\dot{T}_- -
  (\bar{P}[R])^{\mbox{.}},
\eeann
where we have used the well-known formula $[XY] = \bar{X}[Y] + \bar{Y}[X]$,
valid for any two functions $X$ and $Y$. 
 
The terms
\[
  b_\pm\bar{R}\sqrt{\left|1 - \frac{2E_\pm}{\bar{R}}\right|}\,
  \mbox{sh}_{a_\pm}\frac{P_\pm}{\bar{R}}
\]
that result in the super-Hamiltonian after the substitution (\ref{Rbar}) can
be replaced by
\[
  b_\pm R_\pm\sqrt{|F_\pm|}\,\mbox{sh}_{a_\pm}\frac{P_\pm}{R_\pm},
\]
where
\[
  F_\pm = 1 - \frac{2E_\pm}{R_\pm}.
\]
Indeed, $R_\pm = \bar{R} \mp \pi/2$,
so the replacement amounts to using the constraint $\pi = 0$ in the
super-Hamiltonian; such a procedure does not change the equations of motion
because it preserves the constraint surface (cf.\ Refs.\ \cite{H-T} and
\cite{margu}). Finally, we arrive at the action 
\be
  I^f_s = \int dt\left(P_+\dot{R}_+ - E_+\dot{T}_+ - P_-\dot{R}_-
    + E_-\dot{T}_- + \bar{\nu}[R] - \nu C^f_s \right),
\label{shell-actf}
\ee
where
\be
  C^f_s = b_+R_+\sqrt{|F_+|}\ \mbox{sh}_{a_+}\frac{P_+}{R_+}
    - b_-R_-\sqrt{|F_-|}\ \mbox{sh}_{a_-}\frac{P_-}{R_-} +
      M(\bar{R}).
\label{shell-Hf}
\ee
The Liouville part in the action (\ref{shell-actf}) is split up into two
pieces, each being of the form $P\dot{R} - E\dot{T}$, where $T$ and $R$ are
coordinates in a spacetime---the Schwarzschild spacetime left or right to the
shell---and $P$ and $-E$ are the conjugate momenta. This enables us to
generate transformations of the coordinates on the phase space from
transformations of coordinates $(T,R)$ on the Schwarzschild spacetime.

We observe first that the transformation from the
Schwarzschild coordinates $(T,R)$ to the Eddington-Finkelstein coordinates
$(U,R)$ or $(V,R)$ can be completed to a canonical transformation. This is not
so trivial as it may seem: the problem is that the transformation of the
coordinates contains the momentum $-E$. The dependence on $E$ is harmless for
the Eddington-Finkelstein transformation; it is more serious for the
transformation to the Kruskal coordinates.

\subsection{Eddington-Finkelstein coordinates}
\label{sec:EF}
Let us study the Eddington-Finkelstein transformations. As these
transformations do not change the coordinate $R$, it is not necessary to
distinguish $R_+$ from $R_-$ if we are performing it. Thus, we can substitute
$\pi = 0$ everywhere into the action (\ref{shell-actf}): $R_+ = R_- = R$ and
$\bar{R} = R$. In this way, we return to the action (\ref{shell-act}). In the
following formulas, we shall also suppress the annoyig indices $\pm$.

The first Eddington-Finkelstein transformation, in each quadrant and on
each side of the shell, is given by
\beann
 R_U & = & R, \\
 U & = & T - R - 2E\ln\left|\frac{R}{2E} - 1\right|;
\eeann
a suitable ansatz for the new momenta $P_{UR}$ and $P_U$ is
\bea
 P_{UR} & = & P + R\ln\sqrt{|F|}, 
\label{PU} \\
 P_U & = & P_T = - E. \nn
\eea
A similar ansatz for the second transformation is:
\bea
 R_V & = & R, \nn \\
 V & = & T + R + 2E\ln\left|\frac{R}{2E} - 1\right|, \nn \\
 P_{VR} & = & P - R\ln\sqrt{|F|}, 
\label{PV} \\
 P_V & = & P_T = - E. \nn
\eea
To show that the transformations are canonical, we calculate $dU$ and $dV$,
\beann
 dU & = & dT - \frac{R}{R-2E}dR - 2\left(\ln\left|\frac{R}{2E} - 1\right| -
   \frac{R}{R-2E}\right) dE, \\
 dV & = & dT + \frac{R}{R-2E}dR + 2\left(\ln\left|\frac{R}{2E} - 1\right| -
   \frac{R}{R-2E}\right) dE,
\eeann
and substitute this into the Liouville form. We obtain
\[
  P_{UR}dR + P_U dU - P dR - P_T dT = dG,
\]
where
\[
  G = E^2\ln\left|\frac{R}{2E} - 1\right| + \frac{RE}{2} +
  \frac{R^2}{2}\ln\sqrt{|F|}. 
\]
Similarly,
\[
  P_{VR}dR + P_V dV - P dR - P_T dT = - dG.
\]

The transformation of the super-Hamiltonian $C_s$ depends on the
transformation of the term
\[
  bR\sqrt{|F|}\ \mbox{sh}_a\frac{P}{R}.
\]
We obtain in each  of the four quadrants that
\[
  bR\sqrt{|F|}\ \mbox{sh}_a\frac{P}{R} =
  \frac{bR}{2}\left(\mbox{e}^{\frac{P_U}{R}} +
  F\mbox{e}^{-\frac{P_U}{R}}\right) 
\]
for the transformation to the $U$-charts (we have left out the indices $U$ and
$V$ at $R$). From the definition of $b$ in Table \ref{tab:ab}, we can see that
$b$ is continuous inside each $U$-chart $U_I$ and $U_{II}$. Let us define the
sign $b_U$ by $b_U := b$ so that
\[
  b_U = +1 \quad \mbox{in}\quad U_I := \overline{Q_I \cup Q_{IV}} \setminus
  H^+, 
\]
and
\[
  b_U = -1 \quad \mbox{in}\quad U_{II} := \overline{Q_{II} \cup Q_{III}}
  \setminus H^+.
\]
At the future horizons $H^+$, where $T = +\infty$, $U \rightarrow +\infty$ and
$P_U \rightarrow -\infty$ in such a way that $F\exp(-P_U/R)$ is smooth.

The transformation to the $V$-charts $V_I$ and $V_{II}$ is analogous:
\[
  bR\sqrt{|F|}\ \mbox{sh}_a\frac{P}{R} =
  \frac{abR}{2}\left(\mbox{e}^{-\frac{P_V}{R}} +
  F\mbox{e}^{\frac{P_V}{R}}\right).
\]
We define $b_V := ab$ so that  we have
\[
  b_V = +1 \quad \mbox{in}\quad V_I := \overline{Q_I \cup Q_{III}} \setminus
  H^-,
\]
and
\[
  b_V = -1 \quad \mbox{in}\quad V_{II} := \overline{Q_{II} \cup Q_{IV}}
  \setminus H^-.
\]
Again, the super-Hamiltonian has continuous extension to each $V$-chart. At
the past horizon $H^-$, where $T = -\infty$, $V \rightarrow -\infty$ and $P_V
\rightarrow +\infty$ in such a way that $F\exp(P_V/R)$ is smooth.

The result of this section can be interpreted as a new, connected, phase space
that is covered by 16 charts which overlap and that contains all of the 16
disjoint sectors of the old phase space; the super-Hamiltonian has a
continuous extension to the new phase space. The origins of the Kruskal
manifolds remain excluded, however.

\subsection{Kruskal coordinates}
The Kruskal coordinates $u$ and $v$ are regular everywhere inside the Kruskal
manifold (but they are `singular' at the infinity). Thus, they are suitable to
cover the missing points where the horizons intersect.

In each quadrant, the transformation between the Schwarzschild coordinates
$(T,R)$ and the Kruskal coordinates $(u,v)$ is given by (see, eg.\ Ref.\ 
\cite{M-T-W}): 
\bea
  \frac{R}{2E} & = & K(-uv),
\label{Ruv} \\
  \frac{T}{2E} & = & \ln\left|\frac{v}{u}\right|,
\label{Tuv}
\eea
where the function $K : (-1,\infty) \mapsto (0,\infty)$ is a smooth bijection
defined by its inverse
\be
  K^{-1}(x) = (x - 1)\mbox{e}^x,
\label{K}
\ee
and where the signs of the Kruskal coordinates are defined to be
\[
  \begin{array}{llllll}
   u<0 & \mbox{in} & Q_I\cup Q_{IV}, & u>0 & \mbox{in} & Q_{II}\cup Q_{III},\\
   v<0 & \mbox{in} & Q_{II}\cup Q_{IV}, & v>0 & \mbox{in} & Q_{I}\cup
   Q_{III}. 
  \end{array}
\]
 
To begin with, we derive some useful relations. Eq.\ (\ref{Ruv}) implies:
\be
  F = \frac{K - 1}{K}
\label{FK}
\ee
(we leave out the argument of $K$; it will always be $-uv$). Eqs.\ (\ref{Ruv})
and (\ref{K}) imply:
\[
  -uv = K^{-1}\left(\frac{R}{2E}\right) = \left(\frac{R}{2E} -
  1\right)\mbox{e}^{\frac{R}{2E}} = F\frac{R}{2E}\mbox{e}^{\frac{R}{2E}},
\]
or
\be
  F = - \frac{uv}{K\mbox{e}^{K}}.
\label{Fuv}
\ee
Combining Eqs.\ (\ref{FK}) and (\ref{Fuv}), we obtain that
\be
  K - 1 = - \mbox{e}^{-K}uv.
\label{Kuv}
\ee

The next step is to find a smooth `momentum' to replace $P_\pm$. We know from
the previous subsection that $P_{UR}$ is smooth at the past horizon $H^-$,
where $v = 0$, and $P_{VR}$ at the future horizon $H^+$, where $u = 0$. Eqs.\
(\ref{PU}) and (\ref{Fuv}) give
\[
  P_{UR} = P + \frac{R}{2}\ln|v| + \mbox{smooth at}\ H^-,
\]
and, analogously
\[
  P_{VR} = P - \frac{R}{2}\ln|u| + \mbox{smooth at}\ H^+.
\]
Accordindly, the function $\tilde{P}$ defined by
\[
  \tilde{P} := P + \frac{R}{2}\ln\left|\frac{v}{u}\right|
\]
might be smooth everywhere. This suggests that we try the folowing
transformation of momenta
\bea
  P & = & \tilde{P} - \tilde{E}K\ln\left|\frac{v}{u}\right|,
\label{Ptilde} \\
  E & = & \tilde{E},
\eea
and check whether or not the symplectic form expressed by means of the
variables $u$, $v$, $\tilde{P}$ and $\tilde{E}$ is regular everywhere (from
now on, we shall leave out the tilde over $E$). Recall that all equations are
written without the indices $\pm$; in fact, Eq.\ (\ref{Ptilde}) reads, if
written properly, as follows:
\[
  P_\pm = \tilde{P}_\pm -
  \tilde{E}_\pm K_\pm\ln\left|\frac{v_\pm}{u_\pm}\right|, 
\]
where $K_\pm = K(-u_\pm v_\pm)$, etc.

Let us transform the action to the variables
$(u_\pm,v_\pm,\tilde{P}_\pm,E_\pm)$. Eqs.\ (\ref{Ruv}) and (\ref{Tuv}) yield:
\beann
  dR & = & 2KdE - 2EK'(vdu + udv), \\
  dT & = & 2\ln\left|\frac{v}{u}\right|dE + 2E\left(-\frac{du}{u} +
    \frac{dv}{v}\right). 
\eeann
This together with Eq.\ (\ref{Ptilde}) implies:
\beann
  \lefteqn{PdR - EdT = \tilde{P}dR -
  2E(K^2 + 1)\ln\left|\frac{v}{u}\right|dE} \\
  && \mbox{} + E^2\left(2KK'v\ln\left|\frac{v}{u}\right| +
    \frac{2}{u}\right)du +  
  E^2\left(2KK'u\ln\left|\frac{v}{u}\right| - \frac{2}{v}\right)dv.
\eeann
The first term on the R. H. S. is smooth and the rest is singular. To
get rid of it, we observe that
\beann
  \lefteqn{d\left(-E^2(K^2 + 1)\ln\left|\frac{v}{u}\right|\right) =} \\
  && - 2E(K^2 + 1)\ln\left|\frac{v}{u}\right|dE  + 
  E^2\left(2KK'v\ln\left|\frac{v}{u}\right| + \frac{K^2 + 1}{u}\right)du\\
  && \mbox{} + E^2\left(2KK'u\ln\left|\frac{v}{u}\right| - \frac{K^2 +
      1}{v}\right)dv. 
\eeann
This identity implies:
\beann
  \lefteqn{PdR - EdT = - 2EKd\tilde{P} + E^2(K^2 - 1)\left(-\frac{du}{u} +
      \frac{dv}{v}\right)} \\
  &&  \mbox{} + d\left(-E^2(K^2 +
  1)\ln\left|\frac{v}{u}\right| + 2EK\tilde{P}\right). 
\eeann
The second term on the R. H. S. is regular; indeed, Eq.\ (\ref{Kuv}) gives
\[
  E^2(K^2 - 1)\left(-\frac{du}{u} + \frac{dv}{v}\right) = E^2(K +
  1)\mbox{e}^{-K}(vdu - udv).
\]
Hence, finally, the Liouville form becomes
\bea
  \lefteqn{\left[PdR - EdT\right] = \left[- 2EKd\tilde{P} + E^2(K +
      1)\mbox{e}^{-K}(vdu - udv)\right]} \nn \\ 
  && \mbox{} +  d\left[-E^2(K^2 + 1)\ln\left|\frac{v}{u}\right| +
    2EK\tilde{P}\right]. 
\label{liouv-f}
\eea
The singular part is contained entirely within the last term, which can be
discarded without changing the symplectic structure. We postpone the study of
the resulting symplectic structure to the next section.

The last non-trivial step in the transformation of the action is to transform
the term 
\[
  bR\sqrt{|F|}\ \mbox{sh}_a\frac{P}{R}
\]
in the super-Hamiltonian (\ref{shell-Hf}) (the indices $\pm$ are again left
out). Using Eqs.\ (\ref{Ruv}), (\ref{Fuv}) and (\ref{Ptilde}), we obtain
\[
  bR\sqrt{|F|}\ \mbox{sh}_a\frac{P}{R} = bE\sqrt{K}\mbox{e}^{-K/2}
  \sqrt{|uv|}\ \mbox{sh}_a\left(\frac{\tilde{P}}{2EK} -
  \ln\sqrt{\left|\frac{v}{u}\right|}\right). 
\] 
Eqs.\ (\ref{sh-id}) and (\ref{sh-def}) lead then to
\[
  bR\sqrt{|F|}\ \mbox{sh}_a\frac{P}{R} = E\sqrt{K}\mbox{e}^{-K/2}
  \left(b|u|\exp\left(\frac{\tilde{P}}{2EK}\right) +
  ab|v|\exp\left(-\frac{\tilde{P}}{2EK}\right)\right).
\]
The signs of the Kruskal coordinates as defined at the beginning of this
subsection combine with Table \ref{tab:ab} giving that $b|u| = -u$ and $ab|v| =
v$ in each quadrant. Thus, we arrive at the expression:
\be
  bR\sqrt{|F|}\ \mbox{sh}_a\frac{P}{R} = E\sqrt{K}\mbox{e}^{-K/2}
  \left(-u\exp\left(\frac{\tilde{P}}{2EK}\right) +
  v\exp\left(-\frac{\tilde{P}}{2EK}\right)\right).
\label{Hterm}
\ee

Collecting the results (\ref{liouv-f}) and (\ref{Hterm}), we obtain the final
form of the action:
\be
  I^K_s = \int dt\left([\mbox{} - 2EK\dot{\tilde{P}} + E^2(K +
    1)\mbox{e}^{-K}(v\dot{u} - u\dot{v})] + 2\bar{\nu}[EK] -
    \nu C^K_s \right),
\label{shell-actK}
\ee
where
\bea
  C^K_s & = & \left[E\sqrt{K}\mbox{e}^{-K/2} 
  \left(-u\exp\left(\frac{\tilde{P}}{2EK}\right) +
  v\exp\left(-\frac{\tilde{P}}{2EK}\right)\right)\right] \nn \\ 
  & + & M(E_+K_+ + E_-K_-).
\label{shell-HK}
\eea 
Let us recall that $[X] = X_+ - X_-$ and that $K_\pm = K(-u_\pm v_\pm)$
etc. The action (\ref{shell-actK}) as well as all variables on which it
depends, are smooth everywhere in the new phase space. This phase space is
covered by the coordinates $u_\pm$, $v_\pm$, $\tilde{P}_\pm$ and $E_\pm$ with
ranges $u_\pm \in (-\infty, \infty)$, $v_\pm \in (-\infty, \infty)$,
$\tilde{P}_\pm \in (-\infty, \infty)$ and $E_\pm \in (0, \infty)$; it is the
maximal extension of the old phase space. The super-Hamiltonian
(\ref{shell-Hf}) as well as the function $[R] = 2[EK]$ have continuous
extensions to the new phase space.

\subsection{The symplectic form and Poisson brackets}
In this subsection, we investigate the properties of the symplectic structure
defined by the Liouville form (\ref{liouv-f}). 

Taking the external derivative of the form (\ref{liouv-f}), we obtain
\bea
  \Omega(\delta{\mathbf X},\dot{\mathbf X}) & = & \left[ - 2K(\delta
  E\,\dot{\tilde{P}} - \delta\tilde{P}\,\dot{E}) - 4E^2\mbox{e}^{-K}(\delta
  u\,\dot{v} - \delta v\,\dot{u}) \right. \nn \\
  & + & 2E(K + 1)\mbox{e}^{-K}(v\delta E\,\dot{u} - v\delta u\,\dot{E} -
  u\delta E\,\dot{v} + u\delta v\,\dot{E}) \nn \\
  & - & \left. 2EK'(v\delta\tilde{P}\,\dot{u} - v\delta u\,\dot{\tilde{P}} +
  u\delta\tilde{P}\,\dot{v} - u\delta v\,\dot{\tilde{P}}) \right].
\label{omega-f}
\eea
In the calculation, we have used Eq.\ (\ref{Kuv}) and the identity
\be
  K' = \frac{1}{K\mbox{e}^K},
\label{K-prime}
\ee
which follows from the definition (\ref{K}) of $K$.

The form $\Omega$ must be
non-degenerate in order to define a symplectic structure. The calculation of
the determinant of the corresponding matrix $\Omega_m$ can be simplified by
writing it in the $2\times 2$ block form
\be
  \Omega_m = \left(\begin{array}{cc}
                   \mbox{A} & \mbox{C} \\
                   -\mbox{C}^\top & \mbox{B}
                   \end{array} \right).
\label{o-block}
\ee
Multiplying the second double row by the matrix $-\mbox{CB}^{-1}$ and
adding the result to the first double row, one can see immediately that
\[
  \det \Omega_m = \det(\mbox{A} + \mbox{C}\mbox{B}^{-1}\mbox{C}^\top)\det{B}.
\]
After some easy calculation, this leads to
\[
 \det\Omega_m = \left(\frac{8E^2}{K\mbox{e}^K}\right)^2_-
 \left(\frac{8E^2}{K\mbox{e}^K}\right)^2_+ = (8E^2K')^2_-(8E^2K')^2_+.
\]
The determinant is non-zero at all points of all Kruskal spacetimes.

The block form (\ref{o-block}) helps also to fasten the calculation of the
matrix $\Omega_m^{-1}$, which defines the Poisson brackets\footnote{I shorten
  the subsequent exposition, because most people today may prefer a Maple or
  Mathematica routine.}.  We look for $\Omega^{-1}_m$ in the form
\[
  \Omega^{-1}_m = \left(\begin{array}{cc}
                    \mbox{U} & \mbox{W} \\
                    -\mbox{W}^\top & \mbox{V}
                    \end{array} \right)
\]
and observe that the matrices A, B, U and V all must be proportional to the
matrix 
\[
  \epsilon := \left(\begin{array}{rr}
                    0 & -1 \\
                    1 & 0
                    \end{array} \right).
\]
The equation $\Omega_m\Omega^{-1}_m = 1$ now decomposes into four
equations: 
\be
  \mbox{AU} - \mbox{CW}^\top = 1,\quad \mbox{AW} - \mbox{CV} = 0,
\label{oo-1}
\ee
\be
  -\mbox{C}^\top\mbox{U} - \mbox{BW}^\top = 0,\quad -\mbox{C}^\top\mbox{W} +
  \mbox{BV} = 1. 
\label{oo-2}
\ee
From the second Eq.\ of (\ref{oo-1}) and the well-known identities for
$\epsilon$,
\be
  \epsilon^2 = -1,\quad \epsilon\mbox{M}^\top\epsilon = -
  (\det\mbox{M})\,\mbox{M}^{-1}, 
\label{id-epsilon}
\ee 
valid for all $2 \times 2$ matrices M, we find that $W$ is proportional to
$C^{\top -1}$. The rest of Eqs.\ (\ref{oo-1}) and (\ref{oo-2}) contains only
two independent linear equations, which determine U and V. A straightforward
calculation using Eqs.\ (\ref{K-prime}) and (\ref{id-epsilon}) then leads to
the result  
\be
  \Omega_m^{-1} = \left(\begin{array}{cccc}
  0 & \frac{K}{2} & -\frac{u}{4E} & \frac{v}{4E} \\[.1cm]
    & 0 & \frac{K(K+1)u}{4E} & \frac{K(K+1)v}{4E} \\[.1cm]
    & & 0 & \frac{K\mbox{\footnotesize e}^K}{4E^2} \\
    & & & 0 
  \end{array} \right)
\label{o-invers}
\ee
(the order of the coordinates is $E,\tilde{P},u,v$).

With the help of Eq.\ (\ref{o-invers}), we can study Poisson brackets. We
observe first that
\be
  \{u,v\} = \frac{K\mbox{e}^K}{4E^2} \neq 0,
\label{PBuv}
\ee
and, second, that
\be
  \{u,E\}_{u=v=0} = 0,\quad \{v,E\}_{u=v=0} = 0.
\label{PB-E}
\ee
This has interesting consequences. First, there is no Darboux chart such that
$u$ and $v$ would be two of the corresponding coordinates. Second, a stronger
version of inequality (\ref{PBuv}) can be proved. Consider an arbitrary pair
$(x,y)$ of coordinates in a neighbourhood of the horizon crossing $u=v=0$. If
$x$ and $y$ are to be independent, they must satisfy
\be
  \frac{\partial x}{\partial u}\frac{\partial y}{\partial v} - \frac{\partial
  x}{\partial v}\frac{\partial y}{\partial u} \neq 0.
\label{indep}
\ee
If they are to be coordinates on the spacetime where the shell moves, they
must be independent on $\tilde{P}$:
\[
  x = x(u,v,E),\quad y = y(u,v,E).
\]
Let us calculate the Poisson bracket of the two coordinates at the horizon
crossing. If we take Eqs.\ (\ref{PB-E}) into account, we obtain:
\[
  \left. \{x,y\}|_{u=v=0} = \frac{K\mbox{e}^K}{4E^2}\left(\frac{\partial
  x}{\partial u}\frac{\partial y}{\partial v} - \frac{\partial 
  x}{\partial v}\frac{\partial y}{\partial u}\right)\right|_{u=v=0}.
\]
Eqs.\ (\ref{PBuv}) and (\ref{indep}) then imply that this expression is
non-zero. We have shown the following theorem:
\begin{thm}
Any two independent spacetime coordinates of the shell that are regular at an
intersection of two horizons have a non-zero Poisson bracket with each other
in a neighbourhood of the intersection.
\end{thm}

Can this be interpreted as saying that a spacetime manifold is necessarily
fuzzy near a horizon in the quantum theory? There are at least two caveats.
First, any generic point of any horizon (that is, a point that does not lie at
an intersection of two horizons) has a neighbourhood, where there {\em are}
commuting coordinates. An example is given in the Sec.\ \ref{sec:EF}. Second,
the way from the classical to a quantum theory is longer that it may seem: we
had also to define observables, and the observables must satisfy some
requirements. For example, their classical counterparts are to have vanishing
Poisson brackets with the constraints (for a discussion of this point, see
Ref.\ \cite{margu}). The functions $u$ and $v$ as they stand fail to be so. We
can safely conclude that some more work is necessary to understand the
implications of the theorem.

\subsection*{Acknowledgements}
Useful discussions with H.~Friedrich, J.~Kijowski, K.~V.~Kucha\v{r}, J.~Louko
and B.~G.~Schmidt are thankfully acknowledged.  The author is indebted to the
Max-Planck-Institut for Gravitationsphysik, Potsdam and to the Institute for
Theoretical Physics of the University of Utah, Salt Lake City, for the nice
hospitality and support. Thanks go to National Science Foundation grant
PHY9507719 (U. S. A.), to the Tomalla Foundation, Zurich and to the Swiss
Nationalfonds for a partial support.

\end{document}